\documentclass[conference,10pt]{IEEEtran}

\IEEEoverridecommandlockouts

\usepackage{amsmath,amssymb,epsfig,psfrag}
\usepackage{enumerate}
\usepackage{import}

\usepackage{times,subfigure,epsfig}
\usepackage{wrapfig}

\usepackage[latin1]{inputenc}
\usepackage[T1]{fontenc}
\usepackage{rotating}
\usepackage{multirow}
\usepackage{array}
\usepackage{graphicx}
\usepackage{epstopdf}
  \usepackage{amsbsy}
\usepackage{amssymb}
\usepackage{amscd}
\usepackage{amsmath}

\usepackage{cite}
\usepackage{latexsym}
\usepackage{graphicx}
\usepackage{amsmath}
\usepackage{amssymb}
\usepackage{amsfonts}

\def\zf{{{{\textrm{\tiny ZF}}}}}
\def\mr{{{{\textrm{\tiny MR}}}}}
\def\mmf{{{{\textrm{\tiny mm}}}}}
\def\ortho{{{{\textrm{\tiny OA}}}}}
\def\jbb{{{{\textrm{\tiny JBB}}}}}

\def\bU{{\boldsymbol{U}}}

\def\bx{{\boldsymbol{x}}}
\def\bX{{\boldsymbol{X}}}
\def\bn{{\boldsymbol{n}}}

\def\bI{{\boldsymbol{I}}}
\def\bz{{\boldsymbol{z}}}

\def\bzero{{\boldsymbol{0}}}

\def\by{{\boldsymbol{y}}}

\def\bQ{{\boldsymbol{Q}}}

\def\bh{{\boldsymbol{h}}}

\def\bg{{\boldsymbol{g}}}

\def\bG{{\boldsymbol{G}}}
\def\bC{{\boldsymbol{C}}}

\def\bPsi{{\boldsymbol{\Psi}}}
\def\bpsi{{\boldsymbol{\psi}}}

\def\bv{{\boldsymbol{v}}}
\def\bq{{\boldsymbol{q}}}

\newcommand{\pp}[1]{{\left( #1 \right)}}

\newcommand{\tr}[1]{{\mbox{Tr} \left[ #1 \right]}}

\newcommand{\snorm}[1]{{ \left\Vert #1 \right\Vert^2 }}

\newcommand{\E}[1]{{ \mbox{E}\left[ #1 \right] }}

\usepackage{tikz}
\usetikzlibrary{decorations}
\usetikzlibrary{decorations.pathmorphing}
\usetikzlibrary{decorations.text}
\usetikzlibrary{patterns}

\usetikzlibrary{external}
\usepackage{tabu}

 \title{Joint Beamforming and Broadcasting in Massive MIMO}

 \author{Erik G. Larsson and H. Vincent Poor \thanks{E. G. Larsson is
     with the Dept.\ of Electrical Engineering (ISY), Link\"oping
     University, Link\"oping, Sweden. H. V. Poor is with the Dept. of
     Electrical Engineering, Princeton University, Princeton, NJ,
     USA. Parts of this work were performed when the first author was a
     visiting fellow at Princeton University.} \thanks{This work was
     supported in part by the Swedish Research Council (VR), ELLIIT,
     and the U.S. National Science Foundation under Grants CNS-1456793
     and ECCS-1343210.} \thanks{\copyright 2016 IEEE. Personal use
      of this material is permitted. Permission from IEEE must be obtained
       for all other uses, in any current or future media, including
        reprinting/republishing this material for advertising or promotional purposes,
         creating new collective works, for resale or redistribution to servers or lists, 
         or reuse of any copyrighted component of this work in other works.} \thanks{This paper will appear in the IEEE Transactions on Wireless Communications, 2016, DOI: 10.1109/TWC.2016.2515598.}}

 \usetikzlibrary{positioning}

\tikzset{
  mybox/.style = {
  thick,
     minimum height=2cm, 
     inner sep=3pt, 
     draw},
}

\begin{document}
 
\maketitle

\begin{abstract} 
The downlink of a massive MIMO system is considered for the case in
which the base station must concurrently serve two categories of
terminals: one group to which imperfect instantaneous channel state information (CSI)
is available, and one group to which no CSI is available.  Motivating
applications include broadcasting of public channels and control
information in wireless networks.

A new technique is developed and analyzed: joint beamforming and
broadcasting (JBB), by which the base station beamforms to the group
of terminals to which CSI is available, and broadcasts to the other
group of terminals, to which no CSI is available. The broadcast
information does not interfere with the beamforming as it is placed in
the nullspace of the channel matrix collectively seen by the terminals
targeted by the beamforming. JBB is compared to orthogonal access
(OA), by which the base station partitions the time-frequency
resources into two disjunct parts, one for each group of terminals.

It is shown that JBB can substantially outperform OA in terms of required 
total radiated  power for given rate targets.  
\end{abstract}
 
\section{Introduction}
\label{sec:intro}
 
Massive MIMO \cite{Mar2010} is a leading technology candidate for 5G
wireless access.
The main concept is that hundreds of base station antennas act
phase-coherently together and serve tens of terminals in the same
time-frequency resource.  Different base stations, however, do not
cooperate.  A fundamental assumption in massive MIMO is that the base
station antenna array can acquire instantaneous channel state
information (CSI) to the terminals, so that closed-loop beamforming
can be applied.  This is possible by operating in time-division duplex
(TDD) mode, with the base station acquiring CSI from uplink pilots,
and relying on reciprocity of the propagation channel.

In wireless networks, the base station will also need to
broadcast\footnote{The word ``broadcast'' here means transmitting
  common data intended to an unknown number of terminals, and must not
  be confused with the ``broadcast channel'' in information theory.}
information to terminals to which it has no CSI. Practical examples of
when broadcasting is desired in cellular systems include: delivery of
broadcast content \cite{tr027}; evolved multimedia broadcast/multicast
services \cite{lecompte2012}; the transmission of public ``beacon''
channels; and the transmission of user-specific control messages
intended to ``wake up'' a particular terminal and instruct it to send
uplink pilots.

When CSI is unavailable at the base station, beamforming is impossible
and the only way of benefitting from multiple antennas is to use
space-time coding, which does not offer multiplexing or array gains.
Throughout, we call the terminals to which \emph{beamforming} is
performed (using imperfect, instantaneous CSI) ``B-terminals'', and
all \emph{other} terminals in the cell (for which no CSI is available)
``O-terminals''. In general, there is an arbitrary number of O-terminals in the cell.

There are two main ways of accommodating the broadcasting
functionality:
\begin{enumerate}
\item A fraction $\epsilon$ of the available time-frequency resources
  can be set aside for the broadcasting to the O-terminals.  The
  remaining fraction, $1-\epsilon$, of the resources, are then used
  for beamforming to the B-terminals. This approach is termed
  orthogonal access (OA) here.

\item As proposed in preliminary form in \cite{larsson2015spawc} and further developed here, 
the
  base station may concurrently beamform coherently to the B-terminals
  and broadcast to the O-terminals.  This is made possible by placing
  the signals aimed at the O-terminals in the nullspace of the channel
  matrix of the B-terminals. This scheme, called joint beamforming and
  broadcasting (JBB) here, is in turn possible owing to the surplus
  of spatial degrees of freedom in massive MIMO.
\end{enumerate}
This paper analyzes and compares OA and JBB in terms of required
radiated power for given rate targets, taking into account the effects
of channel estimation errors and power control.

\subsection{Related Work}

The need for efficient solutions to broadcasting of public information
in wireless networks using massive MIMO technology has been recognized
before by us \cite{KarlssonL2014} and others
\cite{meng2014constant}. However, no known papers address the specific
problem at hand. Remotely related, reference \cite{6612121} proposed
schemes for multicasting to a known set of terminals for which
imperfect instantaneous CSI is available. Multicasting with
per-antenna power constraints was introduced in
\cite{Christopoulos2014}, and specifically for large antenna arrays in
\cite{Christopoulos2015}.  Reference \cite{joudeh2015sum} considered
combined broadcast/multicast transmission of common and private
symbols, which is a different problem.

JBB exploits the surplus of spatial degrees of freedom in  massive MIMO
systems. In this context, it is  worth pointing out that there are also other possible
uses of these excess degrees of freedom: notably, to achieve  
secrecy by transmitting artificial noise into the channel nullspace \cite{zhu2014secure,zhu2014linear};
to produce per-antenna waveforms with reduced peak-to-average ratios \cite{MoL2013,StL2013,6843875}; and
to suppress out-of-cell interference \cite{DBLP:journals/corr/BjornsonLD14}. 

Rigorous capacity bounds for massive MIMO beamforming performance are
available in the literature: \cite{yang2013performance} for the
downlink, and \cite{NLM2013} for the uplink, most notably. Some of our
analysis uses techniques and results from these references. However,
none of these references dealt with the problem of joint beamforming
and broadcasting.

\section{Preliminaries: Massive MIMO Beamforming}

We consider a single cell comprising a base station with an array of
$M$ antennas, that serves $K$ single-antenna B-terminals; $K<M$. Let
$\bg_k$ be an $M$-vector that represents the channel response, from
the array to the $k$th B-terminal, in a given coherence
interval. ``Coherence interval'' here means the time-frequency space
over which the channel is substantially static. We denote by $\tau_c$
the length (in samples) of a coherence interval.

In the downlink, at time $t$ (``time'' here means sample index in a
given coherence interval), the base station transmits the $M$-vector
\begin{align}\label{eq:1}
\bx(t) =  \sqrt{\rho_b}\cdot {\sum_{k=1}^K \bv_k s_k(t)},
\end{align}
where $\{\bv_k\}$ are beamforming vectors associated with the $K$
terminals, $\{s_k(t)\}$ are symbols aimed at the $K$ terminals at time
instant $t$, and $\rho_b$ is the downlink power.  The symbols
$\{s_k(t)\}$ are assumed to have zero means and unit variances.  The
beamforming vectors $\{\bv_k\}$ are functions of estimates of the
channel responses $\{\bg_k\}$, and normalized such
that\footnote{Throughout this paper, all powers are defined as
  averages over all sources of randomness ($\hat \bG$ in this
  particular equation, since $\{\bv_k\}$ depend on $\hat \bG$).  This
  convention is common in the massive MIMO literature. The reason is
  mostly mathematical convenience.  In principle, somewhat increased
  performance could be obtained by defining a short-term measure of
  power and allocating powers between the coherence
  intervals. However, in massive MIMO, the gain of doing so is not
  appreciable in typical cases because by virtue of the channel
  hardening, $||{\hat\bG}||^2$ fluctuates only slightly from one
  coherence interval to the next.}
\begin{align}\label{eq:vnorm}
\E{ \snorm{ \sum_{k=1}^K \bv_k s_k(t) }} = \E{ \sum_{k=1}^K \snorm{  \bv_k }} =1.
\end{align}
Operationally the beamforming in (\ref{eq:1}) makes sure that power
emitted by the base station array is focused onto the terminals.

The $k$th B-terminal sees an effective scalar channel
  with gain $\bg_k^H \bv_k$. In this paper, we assume that no pilots
  are transmitted on the downlink, and that the B-terminal detects the
  downlink data coherently by assuming that the gain $\bg_k^H \bv_k$
  is equal to its expected value $E[\bg_k^H \bv_k]$.  This assumption
  can be justified thanks to channel hardening: by the law of large
  numbers, {${\bg_k^H \bv_k \approx \E{\bg_k^H \bv_k}}$}.  In
  performance analysis, the effect of the gain error $\bg_k^H \bv_k -
  \E{\bg_k^H \bv_k}$ is then treated as additional effective noise.
  This is a common approach in the massive MIMO literature
  \cite{yang2013performance,NLM2013}, but it is not optimal. For
  example, in low-mobility scenarios where the resource cost of
  downlink pilots is negligible, it is known that the transmission of
  downlink pilots improves performance \cite{NLM:13:ACCCC}. Also,
  practical systems may use downlink pilots for various other
  practical reasons; certain downlink reference signals are typically
  transmitted in all wireless systems to enable synchronization and
  acquisition. Finally, we note that it is possible for the terminal
  to obtain a better estimate of $\bg_k^H \bv_k$ than $\E{\bg_k^H
    \bv_k}$ by using blind gain estimation techniques
  \cite{NL:15:ICASSP}.  

By way of contrast, in case no CSI at the base station is available,
then beamforming as in (\ref{eq:1}) is not meaningful. Instead, the
transmitted vectors $\{\bx(t)\}$ may be constructed using space-time
coding.

\section{Joint Beamforming and Broadcasting}\label{sec:jbb}

With joint beamforming and broadcasting (JBB), the base station
simultaneously beamforms to $K$ B-terminals for which it has CSI, and
broadcasts information aimed at the O-terminals.  The fundamental
feature of massive MIMO that makes this possible is that with $M$
antennas and beamforming to $K$ terminals, there are $M-K$ unused
degrees of freedom.  With JBB, the $M-K$ excess degrees of freedom are
exploited by transmitting the broadcast information in a subspace
orthogonal to the channel collectively seen by the $K$ B-terminals.
 
In detail, consider the transmission of $\bx(t)$ on the downlink. The
$k$th B-terminal receives the following at time $t$:
\begin{align}\label{eq:2}
y_k(t) =  \bg_k^H \bx(t)+ w_k(t),
\end{align}
where $w_k(t)$ is noise, assumed to be $CN(0,1)$ here.  Clearly, any
part of the transmitted vector $\bx(t)$ which falls in the nullspace
of the following matrix:
\begin{align}
\bG^H\triangleq [ \bg_1 ,...,\bg_K]^H
\end{align}
will be invisible to all B-terminals. Hence, to $\bx(t)$ formed as in
(\ref{eq:1}), the base station may add any vector that lies in the
nullspace of $\bG^H$.  In particular, the base station may add
broadcasting information aimed at the O-terminals.  Since the base
station does not have CSI to these O-terminals, it cannot beamform to
them. However, it can use space-time coding.

In general, $\bG$ will not be perfectly known at the base station. We
assume that the base station has an estimate $\hat\bG$ of $\bG$.  Let
$\{ \bz(t) \}$ be a sequence of $M$-vectors intended for the
O-terminals.  Instead of (\ref{eq:1}), the base station then transmits
at time $t$ the sum of two terms:\footnote{ Throughout,
  $\Pi^\perp_{\bX}\triangleq \bI-\Pi_{\bX}$, where
  $\Pi_{\bX}\triangleq \bX(\bX^H\bX)^{-1} \bX^H $ denotes the
  projection onto the column space of $\bX$.}
\begin{align}\label{eq:4}
\bx(t) & =  \sqrt{\rho_b}\cdot\pp{  {\sum_{k=1}^K \bv_k s_k(t)} }  +  \sqrt{\rho_o} 
\cdot  {\Pi^\perp_{\hat\bG} \bz(t)},
\end{align}
where $\bz(t)$ is normalized such that
\begin{align}\label{eq:defPsio}
\E{\snorm{ \Pi^\perp_{\hat\bG} \bz(t)}} = 1.
\end{align}
The first term of (\ref{eq:4}) represents data beamformed to the
B-terminals and the second term represents broadcasting information
aimed at the O-terminals.  These two terms are statistically
uncorrelated.  The constants $\rho_b$ and $\rho_o$ represent the
powers spent on the B-terminals and the O-terminals, and
\begin{align}
\rho_d\triangleq \rho_b+\rho_o
\end{align}
is the total downlink power.   

If $\hat\bG$ is an accurate estimate of $\bG$, then
\begin{align}
 \bg_k^H \Pi^\perp_{\hat\bG} \approx \bzero
 \end{align}
for all $k$, so the B-terminals will not see significant interference
arising from signals aimed at the O-terminals.  The O-terminals will,
however, see interference from the beamformed transmission aimed at
the B-terminals. 

\section{Construction of $\bz(t)$}

OA is a special case when some resources are set aside for only
transmission to the O-terminals and on these resources, $\bx(t) =
\sqrt{\rho_o} \cdot \bz(t)$.  Let $\bh$ represent the channel between
the array and an O-terminal.  Both with OA and JBB, the O-terminals
will not know $\bh$ and hence the transmission aimed at the
O-terminals, encoded in $\{\bz(t)\}$, must be noncoherent or include
pilots.  With JBB, an O-terminal will not see the effect of the
projection $\Pi^\perp_{\hat\bG} $ explicitly. Instead, the O-terminal
effectively sees $\bz(t)$ transmitted over a channel with response $
\Pi^\perp_{\hat\bG}\bh$. The vector $\bh$ will be unknown to the
O-terminal anyway, and so will be $ \Pi^\perp_{\hat\bG}\bh$.

Henceforth, we assume that $\bz(t)$ is confined to a subspace of
dimension $M'$, where $M'\le M$. Then we can write
\begin{align}\label{eq:uq}
\bz(t)=\bU\bq(t)
\end{align}
for some $M'$-vector $\bq(t)$ that consists of encoded information to
the O-terminals, where $\bU$ is a semi-unitary $M\times M'$ matrix;
$\bU^H\bU=\bI$. As a possible special case, $M'=M$ and then, we may
take $\bU=\bI$ without loss of generality. As another (albeit
uninteresting) special case, $M'=1$, which corresponds to
``beamforming'' with a channel-independent beamforming vector given by
the sole column of $\bU$.  The matrix $\bU$ is unknown to the
O-terminals.  We discuss some specifics of the choice of $\bU$ later
in this section.

The idea of confining $\bz(t)$ to lie in a low-dimensional subspace
was independently proposed by several authors
\cite{KarlssonL2014,meng2014constant}.  The motivation is that without
this structure $\{\bz(t)\}$ would have to contain $M$ pilot
vectors. If $M$ is comparable to $\tau_c$ then a very large fraction
of the downlink resources would have to be spent on pilots.  This
situation may well arise in massive MIMO: Consider an $M=100$-antenna
array serving a suburban environment using a 2~GHz carrier with 1 ms
coherence time and 200 kHz coherence bandwidth; then $\tau_c=200$. If
$M>\tau_c$, then downlink training would even be impossible.  By
confining $\bz(t)$ to have the form in (\ref{eq:uq}), only $M'$
downlink pilot vectors are needed.  The constant $M'$ can then be
selected such that $M'\ll \tau_c$.

Space-time coding in the $M'$-dimensional subspace offers spatial
diversity of order $M'$.  Therefore, in environments with no frequency
or time diversity, $M'$ should not be too small.  Conversely, if there
is sufficient time and frequency diversity (outer coding over many
coherence intervals), not much performance is lost by confining
$\bz(t)$ to an $M'$-dimensional subspace \cite{KarlssonL2014}.

When $\bz(t)$ is constructed according to (\ref{eq:uq}) then $\bq(t)$,
rather than $\bz(t)$, should be generated by space-time coding.  Here
we will assume that $\bq(t)$ has independent $CN(0,\xi)$ elements,
where $\xi$ is chosen such that (\ref{eq:defPsio}) is satisfied. This
is not necessarily optimal but serves as a sound starting point in
order to analyze the potential of JBB. In practice, some variant of
space-time block coding may be used, as suggested in
\cite{KarlssonL2014}. 

In the case of JBB, we will assume that $\bU$ depends on $\hat\bG$ in
such a way that {${\Pi^\perp_{\hat{\bG}} \bU = \bU}$}. This
assumption is made mainly for analytical convenience.  In practice
this requires $\bU$ to be random and selected anew in each coherence
interval, but this is no restriction as the effective channel seen by
an O-terminal is unknown anyway. This assumption requires that $M'\le
M-K$, otherwise $\bU$ cannot fit into the nullspace of $\hat\bG^H$.

In the case of OA, $\bU$ may be either fixed or selected randomly in
each coherence interval subject to the condition that
$\bU^H\bU=\bI$. There is no restriction on $M'$; it may range from $1$
to $M$.  As far as the choice of $\bU$ is concerned, OA can be handled
as a special case by letting $K=0$ so that $\hat\bG$ is empty and
{${\Pi^\perp_{\hat{\bG}} = \bI}$}.

 Under the assumptions made, 
\begin{align}
\E{ \snorm{\Pi^\perp_{\hat\bG}\bz(t)}}  
  &  =  \E{ \snorm{\Pi^\perp_{\hat\bG}\bU\bq(t)}} \nonumber \\
&  =  \xi\cdot \E{\tr{\bU^H\Pi^\perp_{\hat\bG} \bU}}  \nonumber \\
& = \xi\cdot \E{\tr{\bU^H \bU}} = \xi M' .
\end{align}
Hence, in order for (\ref{eq:defPsio}) to be satisfied, we must have
\begin{align}\label{eq:xi1m}
\xi = \dfrac{1}{M'}.
\end{align}

In independent Rayleigh fading, as we will see in the analysis in
Sections~\ref{sec:JBBperf} and \ref{sec:ortho}, the only assumptions
needed on $\bU$ are that $\bU^H\bU=\bI$ and
{${\Pi^\perp_{\hat{\bG}} \bU = \bU}$}.  In practice, however, in case
some terminals do not experience independent Rayleigh fading, it may
be wise to randomize $\bU$ as much as possible under these given
constraints. To generate such a ``maximally random'' $\bU$, one may
first compute an arbitrary semi-unitary $M\times (M-K)$ matrix $\bQ$ whose columns
span the orthogonal complement of the
column space of $\hat\bG$.  This matrix  $\bQ$ then satisfies $\bQ^H\bQ=\bI$
and $ \bQ\bQ^H=\Pi^\perp_{\hat{\bG}}$. Then, generate an isotropically
distributed \cite{stewart1980efficient} $(M-K)\times (M-K)$ random matrix
$\bPsi$. Finally, let $\bU$ be the $M'$ first columns of $\bQ\bPsi$.

One could also in principle, in case the fading is known to deviate
from independent Rayleigh and the correlation structure is known,
optimize $\bU$ based on the available side information on the
covariance of the O-terminal channels'. More sophisticated schemes
that perform stochastic beamforming and space-time coding \cite{WuMS}
could also be used.  We do not pursue that possibility in this paper
however, as it is unclear to what extent the correlation structure of
the fading can be known.  In particular, some O-terminals may be
silent for a long time so that the base station has no correlation
information to them; also, if there are many O-terminals with
different channel correlation then there is no single one-fits-all
correlation that would be representative for every O-terminal. In
addition, it appears that no clean closed-form performance results
emerge under such assumptions.

\section{Performance  of Joint Beamforming and Broadcasting}\label{sec:JBBperf}

In this section, we derive lower bounds on the capacity for the
B-terminals and O-terminals when JBB is used.  Modified versions of
these formulas apply when OA is used; see
Section~\ref{sec:ortho}. Throughout, we assume that the terminals are
subject to independent Rayleigh fading. That is, $\{ \bg_k\}$ are
independent, and each $\bg_k$ has independent elements with
distribution $CN(0,\beta_k)$ where $\beta_k $ represents the path loss
of the $k$th terminal.

\subsection{Performance for the B-Terminals}

\subsubsection{Channel Estimates}

We assume that estimates of the channels $\{\bg_k\}$ have been
obtained by the base station based on measurements on mutually
orthogonal uplink pilot sequences transmitted by the terminals, as in
\cite{yang2013performance} and \cite{NLM2013}. These pilot sequences
are $\tau_p^u$ symbols long, where $\tau_c \ge \tau_p^u\ge K$.  The
estimate of $\bg_k$, for $k=1,\ldots,K$, can be written as
\begin{align}\label{eq:b3ce}
\hat\bg_k = \bg_k + \tilde\bg_k,
\end{align}
where $\tilde \bg_k$ is the estimation error.  If MMSE estimation is
used, a straightforward calculation shows that $\hat\bg_k$ and $\tilde
\bg_k$ are mutually uncorrelated, zero-mean Gaussian with covariances
\begin{align}
\E{ \hat \bg_k  \hat \bg_k ^H}  = & \gamma_k \bI
\\
\E{
\tilde \bg_k\tilde \bg_k^H} = & \pp{\beta_k - \gamma_k}\bI,
\end{align}
where we defined
\begin{align}
\gamma_k\triangleq \dfrac{\tau_p^u\rho_u\beta_k^2}{1+\tau_p^u\rho_u\beta_k},
\end{align}
and where $\rho_u$ is the uplink SNR, 
defined as the SNR measured at any of the base station antennas if a terminal with
$\beta_k=1$  transmits with unit power.

\subsubsection{Beamforming}\label{sec:Bbeam}

The $k$th B-terminal receives the following at time $t$:
\begin{align}\label{eq:b1}
y_k(t) & =  \sqrt{\rho_b}    \cdot  \bg_k^H \pp{   {\sum_{k'=1}^K \bv_{k'} s_{k'}(t)}   }
\nonumber \\ & +
 \sqrt{\rho_o} \cdot \bg_k^H  {\Pi^\perp_{\hat\bG} \bU\bq(t)}  + w_k(t) 
\end{align}
where $w_k(t)$ is $CN(0,1)$ noise. The beamforming vectors $\{\bv_k\}$
are computed based on estimates of $\{\bg_k\}$ obtained in the
uplink. Henceforth, we consider maximum-ratio (MR) and zero-forcing
(ZF) processing.  For MR,
\begin{align}\label{eq:mr}
\bv_k = \bv_k^\mr  \triangleq \sqrt{\dfrac{\eta_k}{M\gamma_k}} \hat \bg_k,
\end{align}
and for ZF,
\begin{align}\label{eq:zf}
\bv_k = \bv_k^\zf  \triangleq \left[ \sqrt{\eta_k \gamma_k (M-K)} \hat\bG (\hat\bG^H\hat\bG)^{-1} \right]_{:,k}
\end{align}
where $[\cdot]_{:,k}$ denotes the $k$th column of a matrix.
In (\ref{eq:mr}) and (\ref{eq:zf}), $\{\eta_k\}$ are power control parameters that satisfy
\begin{align}\label{eq:sumeta}
\sum_{k=1}^K \eta_k =1.
\end{align}
(We assume that the base station always expends full power.)  With
$\{\eta_k\}$ chosen as in (\ref{eq:sumeta}), $\{\bv_k^\mr\}$ and
$\{\bv_k^\zf\}$ satisfy (\ref{eq:vnorm}).  In massive MIMO, only slow
power control is used so $\{\eta_k\}$ depend only on the path losses
$\{\beta_k\}$.

\subsubsection{Achievable Rate}

No downlink pilots are used, and instead, the B-terminals rely on
channel hardening.  Using (\ref{eq:mr}) and (\ref{eq:zf}) we
can rewrite (\ref{eq:b1}) in terms of a ``useful signal term'' plus a
sequence of mutually uncorrelated noise and interference terms, as
follows.
\begin{itemize}
\item
For MR beamforming:
\begin{align}\label{eq:b3}
y_k(t) &   =   \sqrt{\dfrac{\rho_b\eta_k}{M\gamma_k}}  \cdot \E{\snorm{\hat\bg_k }} s_k(t)  \nonumber\\
& \qquad +    {  \sqrt{\dfrac{\rho_b\eta_k}{M\gamma_k}}  \cdot  \pp{ \snorm{\hat\bg_k }- \E{\snorm{\hat\bg_k }}} s_k(t)}  \nonumber\\ 
& \qquad -    \sqrt{\rho_b}   \cdot  \tilde\bg_k^H  \pp{\sum_{k'=1}^K \bv_{k'}^\mr s_{k'}(t) } \nonumber\\
& \qquad +  \sqrt{ {\rho_b} }  \cdot \hat\bg_{k}^H  \pp{ \sum_{k'=1,k'\neq k}^K  \sqrt{\eta_{k'}}  \bv^\mr_{k'} s_{k'}(t)} \nonumber \\ 
& \qquad +   \sqrt{\rho_o}   \cdot  \bg_k^H  \Pi^\perp_{\hat\bG} \bU\bq(t)  + w_k(t).
\end{align}
The first term in (\ref{eq:b3}) represents the useful signal and is
equal to $s_k(t)$ weighted by a deterministic constant.  The second
term represents the channel gain uncertainty at the terminal.  The
third term stems from channel estimation errors.  The fourth term
(summation of $K-1$ terms) stems from intracell interference.  The
fifth term stems from transmissions aimed at the O-terminals, but
which are partly seen by the $k$th B-terminal since
$\Pi^\perp_{\hat\bG}\neq \Pi^\perp_{\bG}$.  The sixth term is the
thermal noise.  The variances of the first four terms are known from
\cite{yang2013performance} and \cite{NLM2013}. Details are omitted
here. The variance of the fifth term, which is specific to JBB, is
shown in Appendix A to be equal to
  \begin{align}\label{eq:leakagemag}
    \rho_o \cdot \E{ \left| \bg_k^H  \Pi^\perp_{\hat\bG} \bU\bq(t) \right|^2} = \rho_o   (\beta_k-\gamma_k).
  \end{align}
  (The expectation here is with respect to all sources of randomness;
  hence the result is a deterministic constant.)  Hence, using
  arguments in \cite{yang2013performance,NLM2013,HH:03:IT} we have the
  following achievable rate for the $k$th terminal:
\begin{align}\label{eq:Rmr}
R_k^\mr =  \log_2\pp{1 +
  \dfrac{M\rho_b\gamma_k\eta_k}{\rho_b\beta_k+\rho_o(\beta_k-\gamma_k)  + 1}} .
\end{align}

\item For ZF beamforming:
\begin{align}\label{eq:b3a}
y_k(t) & =  
 \sqrt{(M-K)\rho_b\gamma_k\eta_k} s_k(t) \nonumber \\ & \quad - \sqrt{\rho_b}\cdot \tilde\bg_k^H \pp{\sum_{k'=1}^K \bv_{k'}^\zf s_{k'}(t) }  \nonumber 
 \\ 
& \quad +  \sqrt{\rho_o}   \cdot  \bg_k^H  \Pi^\perp_{\hat\bG} \bU\bq(t)  + w_k(t).
\end{align}
Here, the first term represents the desired signal scaled by a
deterministic constant.  The second
term stems from effects of channel estimation errors, the third term
is leakage from the transmission aimed at the O-terminals and the
fourth term is noise. 
The variances of the first two terms are known
\cite{yang2013performance,NLM2013} and the variance of the third term
is the same as in the case of MR beamforming. The achievable rate is thus
\begin{align}\label{eq:rkzf}
R_k^\zf & =  
\log_2\pp{1+\dfrac{(M-K)\rho_b\gamma_k\eta_k}{(\rho_b+\rho_o)(\beta_k-\gamma_k)+1}}.
\end{align}
\end{itemize}

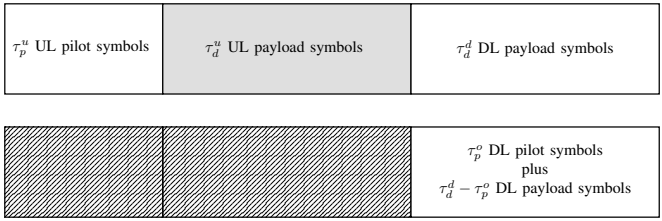
\begin{figure}[t!]
\centerline{
\scalebox{0.6}{
  \begin{tikzpicture}[node distance=-\pgflinewidth,scale=0.3]
\node [mybox=1cm, minimum width=3.5cm,  fill=white, align=center] (c0pos0) { $\tau_p^u$  UL  pilot symbols };
\node [mybox,minimum width=5.5cm, fill=gray!25,right=of c0pos0, align=center] (rxst0) { $\tau_d^u$  UL  payload symbols  };
\node [mybox=1cm,minimum width=5.5cm, fill=white,right=of rxst0, align=center] (c0pos1) {$\tau_d^d$  DL  payload symbols};
  \end{tikzpicture}
}
}

~

\centerline{
\scalebox{0.6}{
  \begin{tikzpicture}[node distance=-\pgflinewidth,scale=0.3]
\node [mybox=1cm, minimum width=3.5cm, pattern=north east lines, align=center] (c0pos0) {  };
\node [mybox,minimum width=5.5cm,right=of c0pos0, pattern=north east lines,align=center] (rxst0) {   };
\node [mybox=1cm,minimum width=5.5cm, fill=white,right=of rxst0, align=center] (c0pos1) { $\tau_p^o$ DL  pilot symbols \\ plus \\ $\tau_d^d-\tau_p^o$  DL  payload symbols };
  \end{tikzpicture}
}
}
\caption{Split of the $\tau_c$ symbols in a coherence interval with
  JBB, from the B-terminal perspective (upper) and the O-terminal
  perspective (lower).\label{fig:cohb}}
\end{figure}

To compute a downlink net sum-spectral efficiency we assume that out
of $\tau_c$ symbols in each coherence interval, $\tau_p^u$ symbols are
used for uplink pilots (as above), $\tau_d^u$ symbols are used for
uplink data and $\tau_d^d$ symbols are used for downlink data, where
the uplink/downlink split is symmetric so that $\tau_d^u=\tau_d^d$; see Figure~\ref{fig:cohb}.
In Figure~\ref{fig:cohb},  $\tau_p^o$ is the number of symbols out of the $\tau_d^d$ long
downlink part of the coherence interval that are set aside for pilots
to the O-terminals; to be explained in Section~\ref{sec:pilotphase}.
The net downlink sum-spectral efficiency in the cell is then
\begin{align}\label{eq:b5}
  R_{\scriptsize\mbox{b,sum-net}} & \triangleq \frac{\tau_d^d}{\tau_c}
  \sum_{k=1}^K R_k \nonumber \\ & = \frac{1}{2}
  \pp{1-\frac{\tau_p^u}{\tau_c}} \sum_{k=1}^K R_k \quad
  \mbox{b/s/Hz/cell},
\end{align}
where $R_k$ is taken from (\ref{eq:Rmr}) for MR and (\ref{eq:rkzf})
for ZF.  Note that we consider TDD operation and hence, to obtain
rates all spectral efficiencies should be multiplied with the full
system bandwidth used for both uplink and downlink.

\subsection{Power Control for the B-Terminals}\label{sec:powercont}

We adopt a max-min fairness power control policy that ensures that all
B-terminals in the cell obtain the same rate.  Such power control is
useful to ensure a uniform   quality-of-service in the cell
\cite{yang2014macro}.  The resulting max-min optimal rate also is a neat 
proxy of the performance for the whole cell, expressed in terms only of the path
loss profile $\{\beta_k\}$. To find the max-min operating point, 
$\{\eta_k\}$ should be selected such that (\ref{eq:sumeta}) holds and
such that $R_k^\mr=\bar R^{\mr,\mmf}$ (for MR) respectively $R_k^\zf=\bar R^{\zf,\mmf}$ (for
ZF) for some maximally large max-min optimal rates $\bar R^{\mr,\mmf}$ and $\bar R^{\zf,\mmf}$ and for all $k$.

For MR, equating   (\ref{eq:Rmr}) to $\bar R^{\mr,\mmf}$ and solving for $\eta_k$ yields 
\begin{align}
\eta_k =   \dfrac{\pp{2^{\bar R^{\mr,\mmf}}-1 }\pp{\rho_b\beta_k+\rho_o(\beta_k-\gamma_k) +1}}{M\rho_b\gamma_k }.
\end{align}
Using the constraint (\ref{eq:sumeta})  we then conclude that
\begin{align}\label{eq:etamr}
\eta_k = \eta_k^\mr \triangleq  \dfrac{  {{\rho_b\beta_k+\rho_o(\beta_k-\gamma_k) + 1 }} }{\gamma_k\cdot {\sum_{k'=1}^K  \dfrac{{\rho_b\beta_{k'}+\rho_o(\beta_{k'}-\gamma_{k'}) + 1 }}{ \gamma_{k'}   }} }.
\end{align}
A similar calculation  for ZF yields
\begin{align}\label{eq:etazf}
\eta_k = \eta_k^\zf \triangleq  \dfrac{(\rho_b+\rho_o)(\beta_k-\gamma_k) + 1 }{\gamma_k\cdot {\sum_{k'=1}^K
 {\dfrac{ (\rho_b+\rho_o)(\beta_{k'}-\gamma_{k'}) + 1 }{\gamma_{k'}} }  }}.
\end{align}
Note that $\{\eta_k^\mr\}$ and $\{\eta_k^\zf\}$ depend on both   $ \rho_b$ and $\rho_o$.
The max-min optimal rates (equal for all terminals in the cell) are, for MR respectively ZF: 
\begin{align}
\bar R^{\mr,\mmf} & =  \log_2\pp{1 + \dfrac{M\rho_b}{ \sum_{k=1}^K  \dfrac{{\rho_b\beta_{k}+\rho_o(\beta_{k}-\gamma_{k})  + 1  }}{ \gamma_{k}   }   }}  \label{eq:mrmmf}, \\
\bar R^{\zf,\mmf} & =   \log_2\pp{1+\dfrac{(M-K)\rho_b }{ \sum_{k=1}^K\dfrac{(\rho_b+\rho_o)(\beta_{k}-\gamma_{k})  + 1 }{\gamma_{k}}   }}. \label{eq:zfmmf}
\end{align}
  
\subsection{Performance for the O-Terminals}\label{sec:perfo}

An O-terminal with channel response $\bh$ will receive the following at time $t$:
\begin{align}\label{eq:ocap}
y_o(t) & =  \sqrt{\rho_o} \cdot     \bh_e^H  \bq(t)   
+  \sqrt{\rho_b}   \cdot \bh^H \pp{ \sum_{k=1}^K   \bv_{k} s_{k}(t) } + w_o(t), 
\end{align}
where 
\begin{align}
\bh_e\triangleq      \bU^H \Pi^\perp_{\hat\bG} \bh = \bU^H\bh \nonumber
\end{align}
represents the effective channel through which the O-terminal sees the
$M'$-dimensional signal $\bq(t)$.  In (\ref{eq:ocap}), the first term
represents the signal of interest, the second term is interference
that stems from the beamformed transmissions, and $w_o(t)$ is
$CN(0,1)$ noise.  
We assume that the O-terminal sees independent Rayleigh fading. Then
\begin{align}
  \bh\sim CN(\bzero,\bC_\bh)
\end{align}
where $\bC_\bh=\beta_o\cdot \bI$ and where $\beta_o$ is the path loss
of the O-terminal. Then, $\bh_e$ is zero-mean with covariance matrix
\begin{align}
  \E{ \bh_e\bh_e^H \big| \hat\bG}
  &  =\beta_o\cdot \bU^H  \bU = \beta_o\cdot \bI \nonumber \\
& =  \E{ \bh_e\bh_e^H }  \triangleq \bC_{\bh_e}.  \label{eq:covbhe}
\end{align}
Recall, that $\bU$ depends on $\hat\bG$ as it is selected to lie in
the nullspace of $\hat\bG^H$.  However, the covariance matrix $\bC_{\bh_e}$ is
independent of $\hat\bG$. Therefore, $\bh_e\sim CN(\bzero,\bC_{\bh_e})$.

\subsubsection{Modified JBB---JBB$'$}

\begin{figure}[t!]
\centerline{
\scalebox{0.6}{
  \begin{tikzpicture}[node distance=-\pgflinewidth,scale=0.3]
\node [mybox=1cm, minimum width=3.5cm,  fill=white,align=center] (c0pos0) { $\tau_p^u$  UL  pilot symbols };
\node [mybox,minimum width=5.5cm, fill=gray!25, right=of c0pos0,align=center] (rxst0) { $\tau_d^u$  UL  payload symbols  };
\node [mybox=1cm,minimum width=5.5cm, fill=white,right=of rxst0, align=center] (c0pos1) { $\tau_p^o$ silent symbols \\ plus \\ $\tau_d^d-\tau_p^o$  DL  payload symbols };
  \end{tikzpicture}
}
}

~

\centerline{
\scalebox{0.6}{
  \begin{tikzpicture}[node distance=-\pgflinewidth,scale=0.3]
\node [mybox=1cm, minimum width=3.5cm, pattern=north east lines,align=center] (c0pos0) {  };
\node [mybox,minimum width=5.5cm, pattern=north east lines,right=of c0pos0, align=center] (rxst0) {   };
\node [mybox=1cm,minimum width=5.5cm, fill=white,right=of rxst0, align=center] (c0pos1) { $\tau_p^o$ DL  pilot symbols \\ plus \\ $\tau_d^d-\tau_p^o$  DL  payload symbols };
  \end{tikzpicture}
}
}
\caption{Split of the $\tau_c$ symbols in a coherence interval with
  JBB$'$, from the B-terminal perspective (upper) and the O-terminal
  perspective (lower).\label{fig:coha}}
\end{figure}
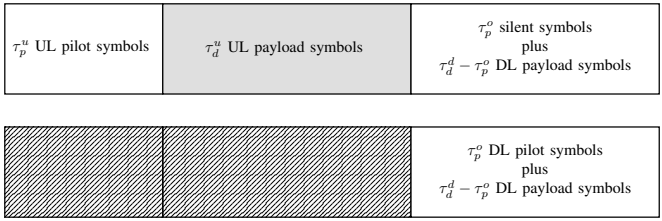

When rigorously analyzing the capacity for the O-terminals, a
technicality arises.\footnote{In preliminary work
  \cite{larsson2015spawc} we took a different approach that avoided
  this technicality. The resulting rate analysis for the O-terminals,
  however, was not entirely rigorous, although numerically it gave
  practically the same result as we derive here.} We will consider a
modified version of JBB where the B-terminals stay silent during the
transmission of pilots to the O-terminals, see Figure~\ref{fig:coha}. We give the name JBB$'$ to
this modified version of JBB, and denote all associated quantities
with $(\cdot)'$. In practice, the original JBB would likely be
preferred over JBB$'$. The only motivation for introducing JBB$'$ is
to facilitate the derivation of an achievable rate without
approximations, as further discussed in Section~\ref{sec:discussion}.

In order to spend the same amount of energy per coherence interval as
with JBB in its original form as described in Section~\ref{sec:jbb},
for JBB$'$, $\rho_b$, must be replaced with
\begin{align}
  \rho'_b \triangleq \frac{\tau_d^d}{\tau_d^d - \tau_p^o}   \cdot \rho_b
  =
  \frac{\frac{1}{2}(\tau_c-\tau_p^u)}{ \frac{1}{2}(\tau_c-\tau_p^u) - \tau_p^o} \cdot \rho_b.
\end{align}
With JBB$'$, the net
downlink B-terminal sum-spectral efficiency is
\begin{align}\label{eq:b51}
  R'_{\scriptsize\mbox{b,sum-net}} & \triangleq
  \frac{\tau_d^d-\tau_p^o}{\tau_c} \sum_{k=1}^K R'_k \nonumber \\
 & =
\frac{1}{2} \pp{1-\frac{\tau_p^u+2\tau_p^o}{\tau_c}} \sum_{k=1}^K R'_k  \quad \mbox{b/s/Hz/cell}.
\end{align}
While $\rho'_b> \rho_b$, the extra loss in degrees of freedom in
(\ref{eq:b51}) renders
$R'_{\scriptsize\mbox{b,sum-net}}<R_{\scriptsize\mbox{b,sum-net}}$ in
general.  On the other hand, the O-terminal performance will be
somewhat better when JBB$'$ is used instead of JBB, since the
O-terminals do not see interference on their pilots.

\subsubsection{Pilot Phase} \label{sec:pilotphase}

The transmission aimed at the O-terminals proceeds in two phases,
first pilots and then payload.

The channel $\bh_e$ is a priori unknown to the O-terminals, and must
be estimated from pilots. Suppose that a string of $\tau_p^o$ downlink
pilot vectors $\{\bq_p(t)\}$ are transmitted to enable the O-terminals
to learn $\bh_e$.  For good performance, these pilots should be
orthonormal. If the energy spent per sample is the same during the
pilot phase and the payload phase, $\{\bq_p(t)\}$ also should satisfy
the power constraint (\ref{eq:defPsio}). Hence, we assume that
\begin{align}\label{eq:dlpilotsdef}
\sum_{t=1}^{\tau_p^o} \bq_p(t)\bq_p^H(t) = \frac{\tau_p^o}{M'} \cdot \bI.
\end{align}
Equation (\ref{eq:dlpilotsdef}) requires that $\tau_c\ge \tau_p^o\ge
M'$.  

Note that in principle, the ratio between the energy per symbol during
the pilot phase and the energy per symbol during the payload phase
could be optimized, but we have not done that here. If $M'\ll\tau_c$,
the pre-log penalty of the pilot transmission is small and for
performance analysis purposes the pilot power can be varied simply by
tuning $\tau_p^o$, subject to $\tau_c\ge \tau_p^o\ge M'$.

An O-terminal receives the $\tau_p^o$ noisy pilot
symbols
\begin{align}\label{eq:yp}
y_o(t) & =   \sqrt{\rho_o} \cdot \bh_e^H \bq_p(t) + w_o(t),
\end{align}
where $w_o(t)$ is $CN(0,1)$ noise.  (Due to the use of JBB$'$ instead
of JBB, there is no interference from the transmission to the
B-terminals here.) The O-terminal correlates $y_o(t)$ with the pilot
sequence to obtain the following statistic:
\begin{align}\label{eq:stat1}
  \by_p \triangleq  \sum_{t=1}^{\tau_p^o} y_o^*(t) \bq_p(t) = \frac{\tau_p^o\sqrt{\rho_o}}{M'} \cdot \bh_e + \bn_p,
\end{align}
where
\begin{align}
  \bn_p \triangleq \sum_{t=1}^{\tau_p^o} w_o^*(t) \bq_p(t)
\end{align}
 has zero mean and covariance
\begin{align}
  \bC_{\bn_p}  & =  \E{ \bn_p\bn_p^H} \nonumber \\
  & =  \E{ \sum_{t=1}^{\tau_p^o}   \sum_{t'=1}^{\tau_p^o}   {w_o^*(t)w_o(t')} \bq_p(t)\bq_p^H(t') } \nonumber \\
&  =
  \frac{  \tau_p^o}{M'} \cdot \bI.
  \end{align}

From $\by_p$, the O-terminal can compute the MMSE estimate of $\bh_e$:
\begin{align}\label{eq:hehat1}
  \hat\bh_e = \E{ \bh_e|\by_p} = \frac{M'\sqrt{\rho_o}\beta_o}{M'+\tau_p^o\rho_o\beta_o} \by_p.
\end{align}
The estimation error $\tilde\bh_e\triangleq \hat\bh_e-\bh_e$ and the
estimate $\hat\bh_e$ are uncorrelated and have covariances
\begin{align}
  \bC_{\tilde\bh_e} & = \E{ \tilde\bh_e\tilde\bh_e^H } = \frac{M'\beta_o}{M'+\tau_p^o\rho_o\beta_o} \cdot\bI,
  \nonumber \\
  \bC_{\hat\bh_e} & = \E{ \hat\bh_e\hat\bh_e^H } = \frac{\tau_p^o\rho_o\beta_o^2}{M'+\tau_p^o\rho_o\beta_o} \cdot\bI
  .
\end{align}
Since all quantities are jointly Gaussian, $\tilde\bh_e$ and $\hat\bh_e$ are independent.

\subsubsection{Payload Phase} 

Next, the O-terminal receives $\tau_d^d-\tau_p^o$ payload symbols. For these
symbols, we have from (\ref{eq:ocap}) that
\begin{align}\label{eq:ocap2}
y_o(t) & =  
 \sqrt{\rho_o} \cdot     \hat \bh_e^H  \bq(t)   -
  \sqrt{\rho_o} \cdot     \tilde \bh_e^H  \bq(t)   \nonumber
\\ & +  \sqrt{\rho'_b}   \cdot   \bh^H \pp{\sum_{k=1}^K  \bv_{k} s_{k}(t)}
 + w_o(t) .
\end{align}
In (\ref{eq:ocap2}), the first term represents the useful signal, the
second term stems from channel estimation errors at the O-terminal,
the third term comprises interference from transmissions aimed at the
B-terminals, and $w_o(t)$ is $CN(0,1)$ noise.

All terms in (\ref{eq:ocap2}) are mutually uncorrelated.  Conditioned
on $\hat\bh_e$, the O-terminal sees the signal $\bq(t)$ transmitted
over a fixed, known channel $\hat\bh_e$, embedded in additive
uncorrelated (non-Gaussian) noise. The distribution of the additive
uncorrelated noise depends on $\hat\bh_e$. However, $\hat\bh_e$ is
known to the O-terminal.  Hence, we must compute the variances of all
terms in (\ref{eq:ocap2}) conditioned on $\hat\bh_e$:
\begin{itemize}
  \item  The conditional received power is
\begin{align}
  & \rho_o\cdot \E{|\hat\bh_e^H\bq(t)|^2 \big| \hat\bh_e} \nonumber \\
  & = \frac{\rho_o}{M'}\cdot \E{||\hat\bh_e||^2 \big| \hat\bh_e} \nonumber \\
  & = \frac{\rho_o}{M'}\cdot ||\hat\bh_e||^2 .
\end{align}

\item Since $\hat\bh_e$ and $\tilde\bh_e$ are independent, the second
  term of (\ref{eq:ocap2}) has conditional variance
\begin{align}\label{eq:v1}
V_1 & \triangleq \rho_o\cdot  \E{|        \tilde \bh_e^H  \bq(t)   |^2 | \hat\bh_e   } \nonumber \\
&  =   \frac{\rho_o}{M'} \cdot \E{||  \tilde \bh_e  ||^2 | \hat\bh_e   } \nonumber \\
&  =   \frac{\rho_o}{M'} \cdot \E{||  \tilde \bh_e  ||^2  } \nonumber \\
&  =  \frac{\rho_o}{M'} \cdot\tr{ \bC_{\tilde\bh_e}}  \nonumber \\
&  =   \frac{M'\rho_o\beta_o}{M'+\tau_p^o\rho_o\beta_o} ,
\end{align}
independently of $\hat\bh_e$.

\item 
  The third term of (\ref{eq:ocap2}) must be handled judiciously, due
  to the interdependence of $\bh$ and $\hat\bh_e$. First note that
  conditioned on $\hat\bG$, $\bU$ is fixed, so from (\ref{eq:stat1})
  and (\ref{eq:hehat1}), $\hat\bh_e$ and $\bh$ are jointly Gaussian with zero means
  and cross-covariance
\begin{align}
  \E{ \bh\hat\bh_e^H  | \hat\bG } =  \frac{\tau_p^o \rho_o \beta_o^2}{M' + \tau_p^o\rho_o\beta_o} \cdot  \bU.
  \end{align}
It follows that (see, e.g., \cite[Lemma~2.4.1]{soderstrom2002discrete})
\begin{align}\label{eq:v2pre}
  \E{\bh\bh^H|\hat\bh_e , \hat\bG } & = \bC_{\bh} -   \E{ \bh\hat\bh_e^H| \hat\bG  }
  \cdot \bC_{\hat\bh_e}^{-1}
  \cdot   \E{ \hat\bh_e \bh^H | \hat\bG } \nonumber \\
  & = \beta_o\cdot\bI - \frac{\tau_p^o\rho_o\beta_o^2}{M'+\tau_p^o\rho_o\beta_o}\cdot\bU\bU^H.
\end{align}
In (\ref{eq:v2pre}) we used that {${\E{\hat\bh_e\hat\bh_e^H|\hat\bG}=\E{\hat\bh_e\hat\bh_e^H } = \bC_{\hat\bh_e}}$}, similarly to in (\ref{eq:covbhe}).
Hence, the third term of (\ref{eq:ocap2})  has  conditional variance
\begin{align}\label{eq:v2}
& V_2   \triangleq \rho'_b\cdot \E{ \left| \sum_{k=1}^K \bh^H \bv_k s_k(t) \right|^2 \Big| \hat\bh_e } \nonumber \\
  & =  \rho'_b\cdot \E{  \sum_{k=1}^K \bv_k^H \bh \bh^H \bv_k \Big| \hat\bh_e } \nonumber \\
  & =  \rho'_b\cdot \E{ \E{  \sum_{k=1}^K \bv_k^H \bh \bh^H \bv_k \Big| \hat\bh_e ,\hat\bG } \Bigg| \hat\bh_e } \nonumber \\
& =   \rho'_b\cdot \E{ \sum_{k=1}^K \bv_k^H \E{ \bh \bh^H \Big| \hat\bh_e,\hat\bG } \bv_k \Big| \hat\bh_e } \nonumber \\
  & =   \rho'_b\cdot \Bigg(\beta_o
\cdot \E{\sum_{k=1}^K ||\bv_k||^2} 
\nonumber \\ &\qquad    - \frac{\tau_p^o\rho_o\beta_o^2}{M'+\tau_p^o\rho_o\beta_o} \cdot \E{ \sum_{k=1}^K \bv_k^H   \bU\bU^H  \bv_k \Big| \hat\bh_e }\Bigg) \nonumber \\
& = \rho'_b\beta_o,
\end{align}
\end{itemize}
independently of $\hat\bh_e$.  In (\ref{eq:v2}) we used
(\ref{eq:vnorm}) and the fact that $\bU^H\bv_k=\bzero$ for all $k$
since $\bU^H\hat\bG=\bzero$ by construction; see (\ref{eq:mr}) and
(\ref{eq:zf}).   We also used that   {${ 
 \E{\sum_{k=1}^K ||\bv_k||^2 \big| \hat\bh_e }   = \E{\sum_{k=1}^K ||\bv_k||^2} }$},
 as  the distribution of $\bU^H\bh$ conditioned on $\bU$ is the same for all $\bU$.
 In (\ref{eq:v2}), when double expectations appear, the
inner expectation is conditioned on $\hat\bG$ and $\hat\bh_e$, and the
outer expectation is with respect to $\hat\bG$ conditioned on
$\hat\bh_e$.

A lower capacity bound is obtained by assuming that the uncorrelated
effective noise in (\ref{eq:ocap2}) is Gaussian.  Averaging over
$\hat\bh_e$ gives the following achievable rate for the O-terminal:
\begin{align}\label{eq:orate1}
R_o & \triangleq   \E{ \log_2\pp{1 + \dfrac{ \frac{\rho_o}{M'} \cdot
      ||\hat\bh_e||^2}{ V_1+V_2+1}}} \nonumber \\
& =  \E{ \log_2\pp{1 + \dfrac{ \frac{\rho_o}{M'} \cdot
      ||\hat\bh_e||^2}{  \frac{M'\rho_o\beta_o}{M'+\tau_p^o\rho_o\beta_o}
      +  \rho'_b\beta_o  +1}}}
\end{align}
In (\ref{eq:orate1}), the expectation is with respect to
$\hat\bh_e$. Since the O-terminal knows $\hat\bh_e$, this average can
be interpreted as an ergodic achievable rate.  This rate only has a
meaning if there is coding across multiple coherence intervals that
see independent fading.

The expectation in (\ref{eq:orate1}) can be calculated
in closed form \cite[Theorem~II.1]{shin}, however, the result
contains exponential integral functions of higher order and is
difficult to interpret intuitively.
To obtain a simple closed-form bound, we use the fact that if $\bpsi$ is  an $M'$-vector   with
  independent $CN(0,1)$ elements, then for any $\alpha>0$,
\begin{align} \label{eq:jensen}
& \E{\log_2\pp{1+\alpha\snorm{\bpsi}}}  \nonumber \\
 & \ge \log_2\pp{1 + \frac{\alpha}{\E{\frac{1}{\snorm{\bpsi}}}}} \nonumber \\
 & = \log_2(1+(M'-1)\alpha).
\end{align}
The first step in (\ref{eq:jensen}) follows from Jensen's inequality and the second step from a 
random matrix theory result \cite[Lemma 2.10]{verdu}.
Since $\hat\bh_e$ has independent Gaussian elements with variance
\begin{align}
  \frac{\tau_p^o\rho_o\beta_o^2}{M'+\tau_p^o\rho_o\beta_o},
  \end{align}
using (\ref{eq:jensen}) on (\ref{eq:orate1}) yields
\begin{align}\label{eq:henormapr}
  R_o & \ge  \log_2\pp{1 + \dfrac{   \frac{M'-1}{M'} \cdot \rho_o\beta_o
    \cdot \frac{\tau_p^o\rho_o\beta_o}{M'+\tau_p^o\rho_o\beta_o} }{   \rho_o\beta_o\cdot \frac{M'}{M'+\tau_p^o\rho_o\beta_o}
      + \rho'_b\beta_o +1    }}  .
\end{align}
The inequality may not be tight if $M'$ is small, but if $M'$ is on the
order of ten, or so, (\ref{eq:henormapr}) should be not only a bound
but also a reasonable approximation.

Taking into account the bandwidth cost of channel training, the net rate for an O-terminal is
\begin{align}\label{eq:orate}
  R_{o,\scriptsize\mbox{net}} \triangleq \frac{\tau_d^d- \tau_p^o}{\tau_c}\cdot R_o
  =
  \frac{1}{2}\pp{1 - \frac{ \tau_p^u + 2\tau_p^o}{\tau_c}} \cdot R_o
\end{align}
 \mbox{b/s/Hz}.

\section{Performance  of Orthogonal Access}\label{sec:ortho}

Next we consider the option of orthogonal access (OA), where
transmissions to the B-terminals and the O-terminals take place on
orthogonal resources. Let $\epsilon$ be the fraction of the available
coherence intervals that are used for transmission to the O-terminals
so that $1-\epsilon$ is the fraction that remains for transmission to
the B-terminals.  Also, let $\rho_b^\ortho$ and $\rho_o^\ortho$ be the
powers spent on the B- respectively O-terminals with OA. Generally, in
what follows, the superscript $(\cdot)^\jbb$ will be used to denote
quantities pertinent to JBB, as derived in previous sections, and the
superscript $(\cdot)^\ortho$ will be used for OA.

\subsection{Performance for the B-Terminals}

The B-terminal rates with max-min fairness power control  
are obtained by setting $\rho_o=0$ and $\rho_b=\rho_b^\ortho$ in
(\ref{eq:mrmmf}) and (\ref{eq:zfmmf}) and weighting the throughput by
$1-\epsilon$:
\begin{align}
\bar R^{\mr,\mmf,\ortho} & = (1-\epsilon)  \log_2\pp{1 + \dfrac{M\rho_b^\ortho}{ \sum_{k=1}^K  \dfrac{{ \rho_b^\ortho\beta_{k} + 1  }}{ \gamma_{k}   } }}  \label{eq:rmrmmoa} \\
\bar R^{\zf,\mmf,\ortho} & = (1-\epsilon)  \log_2\pp{1+\dfrac{(M-K)\rho_b^\ortho }{ \sum_{k=1}^K\dfrac{ \rho_b^\ortho (\beta_{k}-\gamma_{k}) + 1 }{\gamma_{k}} }} . \label{eq:rzfmmoa} 
\end{align}
With OA there is no need for the B-terminals to be silent during the
transmission of pilots to the O-terminals. Hence, the net sum-rates are obtained by multiplying
$ \bar R^{\mr,\mmf,\ortho} $ and $ \bar R^{\zf,\mmf,\ortho} $ with
\begin{align}\label{eq:prelogoa}
\frac{1}{2} \pp{1-\frac{\tau_p^u}{\tau_c}} \cdot K.
\end{align}
similarly to in (\ref{eq:b5}).
Also note that consequently, (\ref{eq:rmrmmoa}) and (\ref{eq:rzfmmoa}) contain $\rho_b$, not $\rho'_b$.

\subsection{Performance for the O-Terminals}

The O-terminal rate is obtained by setting $\rho'_b=0$ in (\ref{eq:orate1})
and weighting by $\epsilon$:
 \begin{align}  \label{eq:Roortho}
R_o^\ortho & = \epsilon \cdot  \E{ \log_2\pp{1 + \dfrac{  \frac{\rho_o^\ortho}{M'}  \cdot \snorm{\hat\bh_e}}{    \frac{M'\rho^\ortho_o\beta_o}{M'+\tau_p^o\rho^\ortho_o\beta_o}
      +1     }}}.
\end{align}
The corresponding bound is, from (\ref{eq:henormapr}):
\begin{align}   \label{eq:Roortho_appr}
R_o^\ortho & \ge \epsilon \cdot \log_2\pp{ 1 + \dfrac{
    \frac{M'-1}{M'}\cdot \rho_o^\ortho\beta_o     \cdot \frac{\tau_p^o\rho^\ortho_o\beta_o}{M'+\tau_p^o\rho^\ortho_o\beta_o} }{
\rho^\ortho_o\beta_o\cdot    \frac{M'}{M'+\tau_p^o\rho^\ortho_o\beta_o} + 1 }} .
\end{align}
Net-rates are obtained by multiplying with
\begin{align}
\frac{1}{2} \pp{1-\frac{\tau_p^u+2\tau_p^o}{\tau_c}} ,
\end{align}
as in (\ref{eq:orate}).

In order to make a fair comparison between JBB$'$ and OA, $\epsilon$
must be chosen such that OA perform at its best.  The find the optimal $\epsilon$ in this respect,
we require that
for a given ``operating point'' in terms of $\rho_b^\jbb$ and
$\rho_o^\jbb$, the corresponding values of $\rho_b^\ortho$ and
$\rho_o^\ortho$ must satisfy
\begin{align}\label{eq:orthopower}
\rho_b^\jbb+\rho_o^\jbb =  (1-\epsilon)\rho_b^\ortho + \epsilon\rho_o^\ortho .
\end{align}
Equation (\ref{eq:orthopower}) guarantees that the total energy spent
in a coherence interval is the same in both cases.  In order for OA to
yield the same B-terminal performance as JBB$'$ does at this operating
point, we require that
\begin{align}
\bar R^{\mr,\mmf,\ortho} & = \bar R^{\mr,\mmf,\jbb'},   \label{eq:mre1} \\
\mbox{respectively}\quad \bar R^{\zf,\mmf,\ortho} & = \bar R^{\zf,\mmf,\jbb'} , \label{eq:zfe1}  
\end{align}
for some $\epsilon$, $0<\epsilon<1$.   Given $\rho_b^\jbb$,
$\rho_o^\jbb$ and $\epsilon$, solving (\ref{eq:mre1}) and
(\ref{eq:zfe1}) for $\rho_b^\ortho$ we can determine how much is the
B-terminal power needed with OA, as follows:
\begin{align}
\rho_b^{\mr,\ortho} & = \dfrac{\pp{2^{\frac{\bar R^{\mr,\mmf,\jbb'}}{1-\epsilon}} -1}
  \sum_{k=1}^K \frac{1}{\gamma_k}}{M -
  \pp{2^{\frac{{\bar R^{\mr,\mmf,\jbb'}}}{1-\epsilon}} -1} \sum_{k=1}^K
  \frac{\beta_k}{\gamma_k}}, \\ 
\rho_b^{\zf,\ortho} & = \dfrac{\pp{2^{\frac{\bar R^{\zf,\mmf,\jbb'}}{1-\epsilon}} -1}
  \sum_{k=1}^K \frac{1}{\gamma_k}}{M - K -
  \pp{2^{\frac{{\bar R^{\zf,\mmf,\jbb'}}}{1-\epsilon}} -1} \sum_{k=1}^K
  \frac{\beta_k-\gamma_k}{\gamma_k}}.
\end{align}
Then, solving (\ref{eq:orthopower}) with respect to $\rho_o^\ortho$,
subject to the constraint that $\rho_o^\ortho\ge 0$, we can find how
much power that remains to spend on the O-terminals.  The solution to
(\ref{eq:orthopower}) may not exist, because of the requirement that
$\rho_o^\ortho\ge 0$.  In case a solution exists, $R_o^\ortho$ is
given by (\ref{eq:Roortho}), and in case no solution exists we set
$R_o^\ortho=0$.  Next, for each operating point we find the value of
$\epsilon$, $0\le \epsilon\le 1$, that maximizes $R_o^\ortho$.  We do
not have a closed-form expression for this optimal $\epsilon$, and in
the numerical examples it was chosen by a grid search from $0$ to
$1$. Typically, performance is not very sensitive to the choice of
$\epsilon$.

Taken together, the above-described procedure gives us, for any
$(\rho_b^\jbb,\rho_o^\jbb)$, the values of
$(\rho_b^\ortho,\rho_o^\ortho)$ for which (\ref{eq:orthopower}) and
(\ref{eq:mre1}) respectively (\ref{eq:zfe1}) hold, and for which
$R_o^\ortho$ is as large as possible.

\section{Discussion}\label{sec:discussion}

The capacity bounds (\ref{eq:mrmmf}) and (\ref{eq:zfmmf}) for the
B-terminal performance, along with the bound (\ref{eq:henormapr}) on
the O-terminal performance, give insights into the impact of the
various system parameters on performance:
\begin{itemize}
\item $M$ and $K$ substantially affect only the performance of the
  B-terminals, but not the performance of the O-terminals.  JBB in
  principle works for any $M$ and $K$ ($K<M$). However, it
  underperforms OA unless $M$ is sufficiently large. This is the
  ``massive MIMO'' aspect of JBB.

\item In terms of B-terminal performance, the leakage that occurs when
  projecting the O-terminal signals onto the nullspace of $\hat\bG^H$,
  rather than that of $\bG^H$, depends only on $\rho_o$ and on the
  quality of the channel state information (as characterized by
  $\gamma_k$).  The better uplink SNR $\rho_u$, the closer is
  $\gamma_k$ to $\beta_k$ and the smaller is this leakage.

\item In terms of O-terminal performance, unless the effects of
  channel estimation errors dominate, the performance is essentially
  determined by $\rho_o$, $\rho'_b$ and $\beta_o$.  Consider
  (\ref{eq:henormapr}).  For the effect of channel estimation
  errors to be negligible, we need
\begin{align}
\tau_p^o \gg \frac{M'}{\rho_o\beta_o}
\end{align}
so the number of downlink pilots must scale with $M'$---consistently
with intuition.

\end{itemize}
A few other technical remarks are in order:
\begin{itemize}
\item For performance analysis, a modification (called JBB$'$) of JBB
  was considered, where the B-terminals are silent during the training
  phase of the O-terminals.  We stress that this modification is not
  necessary, or even desired, if applying JBB in practice. It was only
  introduced in order to enable the calculation of a lower bound on
  ergodic capacity for the O-terminals.
  
  The difficulty with a rigorous analysis of the original JBB scheme
  is, in more detail, the following.  With the original JBB the
  received pilots in (\ref{eq:yp}), will depend on $\hat\bG$ and on
  the (random) symbols transmitted to the B-terminals during the time when
  pilots are transmitted to the O-terminals.  Hence the channel
  estimate $\hat\bh_e$ will also depend on those quantities. This
  dependence must be taken into account when computing the conditional
  (on $\hat\bh_e$) variances in (\ref{eq:v1}) and (\ref{eq:v2}), which
  we were unable to obtain in closed form.

\item Throughout, in order to understand and expose the tradeoffs
  associated with JBB at maximum possible depth, we have focused on a
  single-cell setup.  In a multi-cell setup, additional interference
  will be present from other cells. This interference comprises among
  others so-called ``pilot contamination'' which is known to
  constitute an ultimate limitation in the sense that unlike all other
  interference, it does not go away even if $M\to\infty$
  \cite{Mar2010}.

  Using results known from, for example \cite{yang2013capacity}, one
  can show that the effects of these additional sources of
  interference, when deriving capacity lower bounds for the
  B-terminals, can be accounted for by scaling the numerator and
  augmenting the denominator inside the logarithm in (\ref{eq:Rmr})
  and (\ref{eq:rkzf}) with additional deterministic terms. The rate
  expressions for the O-terminals could also be modified to take into
  account the effects of inter-cell interference. Hence, in principle,
  the analysis here could be extended to a multi-cell setup; however,
  a comprehensive performance evaluation would require serious system
  simulations which in turn requires judicious choices of power
  control policies, pilot reuse and allocation schemes, and
  terminal-base station association algorithms. We believe that such
  simulations could easily obscure the main points we wish to make in
  this paper. Hence, extensions of the performance evaluation to
  multi-cell setups have to be left for future work.
\end{itemize}

\section{Numerical Examples}
     
JBB does not uniformly outperform OA, but there are many situations
when it performs substantially better. Here, we provide some examples
of such cases.  With MR beamforming JBB almost always outperforms
OA. Since JBB is as computationally demanding as ZF, we consider only
ZF beamforming in the examples here. Due to the lack of availability
of performance bounds for JBB, in all comparisons we consider JBB$'$
instead of JBB, even though JBB is expected to perform somewhat better
in practice. However, as in the derivations, we use
$(\rho_b^\jbb,\rho_o^\jbb)$ to define the system operating point.
  
In the numerical examples, $K$ terminals were placed inside an
annulus-shaped cell with outer radius 1 unit and inner radius 0.1
unit.  A standard log-distance path loss model with exponent 4 was
used. However, there was no shadow fading.  Fast fading was modeled as
Rayleigh and independent between the antennas.  The length of the
coherence interval was $\tau_c=500$ symbols, corresponding to mobile
suburban radio access in the 2 GHz-band (2 ms coherence time; 250 kHz
coherence bandwidth).  The uplink cell-edge SNR was $\rho_u=-3$
dB. This SNR corresponds to a gross spectral efficiency of
$\log_2(1+10^{-3/10})\approx 0.6$ b/s/Hz for a reference SISO AWGN
link---however, owing to the large array gain, massive MIMO delivers
good performance even at such low SNRs.

Performance for B-terminals was evaluated in
terms of achievable net sum-rate with max-min power control. 
 Performance for the O-terminals was evaluated in terms of net rate, assuming that the
O-terminals are located at the cell border.
Specifically, 
as functions of the total downlink power $\rho_d^\jbb=\rho_b^\jbb+ \rho_o^\jbb$ and the power ratio
$\rho_o^\jbb/\rho_b^\jbb$, we determine:
\begin{enumerate}[(i)]
\item The set of operating points for which JBB$'$ achieves a
  pre-determined net target sum-rate to the B-terminals of
  $R^\ast_{b,\scriptsize\mbox{sum-net}}$ b/s/Hz---that is, owing to
  the max-min power control, $R^\ast_{b,\scriptsize\mbox{sum-net}}/K$
  b/s/Hz guaranteed to each one of the B-terminals. These are the black curves.

\item The set of operating points for which JBB$'$ delivers a
  predetermined net target rate of $R^\ast_{o,\scriptsize\mbox{net}}$
  b/s/Hz/terminal to the O-terminals. These are the red curves.

\item The set of operating points for which there exist a resource
  split parameter $\epsilon$ and a feasible power allocation
  $(\rho_b^\ortho,\rho_o^\ortho)$ with which OA delivers the same
  B-terminal performance as does JBB$'$, and simultaneously a
  pre-determined O-terminal net target rate of
  $R^\ast_{o,\scriptsize\mbox{net}}$ b/s/Hz/terminal. These are the blue curves.
\end{enumerate}
Figures~\ref{fig:a}--\ref{fig:c} show concrete examples:
\begin{itemize}
\item Figure~\ref{fig:a}: Here, $M=100$ antennas serve a single
  ($K=1$) terminal.  Both the B-terminal and the O-terminals are
  randomly located on the cell border.  The target B-terminal rate is
  $2$ b/s/Hz and the target O-terminal rate is $0.75$
  b/s/Hz.\footnote{Note that while these spectral efficiencies may
    seem low, they are twice as high during the time when transmission
    in the downlink actually takes place.  For comparison with a
    frequency-division duplexing system, all numbers
    should be multiplied by the total bandwidth allocated for both
    uplink and downlink.}  A pilot sequence of length $\tau_p=10$
  symbols was used in the uplink, which is easily afforded given the
  long channel coherence. In the downlink, somewhat arbitrarily,
  $M'=7$ and $\tau_p^o=10$.
  
  The selected operating point can be achieved in two ways: (i) using
  JBB$'$, and (ii) using OA.  These two possibilities correspond to
  the following two intersection points between the curves in the figure: (i)
  when the curve for $2$~b/s/Hz B-terminal performance 
  intersects the curve for $0.75$~b/s/Hz O-terminal performance with
  JBB$'$, and (ii) when the curve for $2$~b/s/Hz B-terminal
  performance  intersects the curve for $0.75$~b/s/Hz
  O-terminal performance with OA.  In terms of required total radiated
  power, JBB$'$ offers savings of about $3$~dB compared to OA.
  
  Note that at the operating point of interest, most
  of the radiated power is spent on the O-terminals: It is expensive
  to reach those terminals since no array gain is available.

\item Figure~\ref{fig:b}: Here, $M=100$ antennas serve $K=10$
  terminals.  The B-terminals were dropped at random in the cell,
  yielding a path loss profile consisting of $K$ values
  $\{\beta_k\}$. The O-terminals are at the cell border, with an
  additional fading margin of 10 dB. This models a scenario in which
  the O-terminals are deeply shadowed and the base station has to expend
  significant resources in order to reach the O-terminals.  The target
  B-terminal rate is $2$ b/s/Hz/terminal ($20$ b/s/Hz sum-rate) and
  the target O-terminal rate is $0.5$ b/s/Hz.  A pilot sequence of
  length $\tau_p^u=30$ symbols is used in the uplink, that is, three
  symbols per terminal, which is afforded without problem given the
  long channel coherence. In the downlink, $M'=7$ and $\tau_p^o=10$.
  The power saving of JBB$'$ compared to OA here is about $2.5$~dB.

\item Figure~\ref{fig:c}: Here, $M=150$ antennas serve $K=30$
  terminals randomly located in the cell. The O-terminals are at the
  cell border (without any extra fading margin).  The B-terminal
  target rate is $1.67$ b/s/Hz/terminal ($50$~b/s/Hz sum-rate) and the
  O-terminal target rate is $0.75$ b/s/Hz.  In the uplink,
  $\tau_p^u=60$ pilot symbols are used and in the downlink, $M'=7$ and
  $\tau_p^o=10$. The gain of JBB$'$ over OA is smaller here, but still
  tangible.

\end{itemize}

Note that the O-terminal rate $R_o$ is a monotonically decreasing
function of the O-terminal path loss $\beta_o$.  This can be seen from
(\ref{eq:henormapr}). Hence, the cell border is the worst possible
location for an O-terminal so in that respect the examples in Figures
\ref{fig:a}--\ref{fig:c} show worst-case performance. In practice, it
could happen that the O-terminals are located closer to the base
station. They could then be served with somewhat higher rate.
However, the increase in rate is marginal in cases of interest.  To
exemplify, Figure~\ref{fig:e1} shows a variation of the result of
Figure~\ref{fig:a}, when the O-terminal is located halfway between the
base station and the cell border. Qualitatively, Figure~\ref{fig:e1}
is similar to Figure~\ref{fig:a}, but a lower total power is required.

To provide additional insight, Table~\ref{table:1} shows for each of
the examples in Figures~\ref{fig:a}--\ref{fig:c} and the two possible
operating points, the following quantities:
\begin{itemize}
\item The optimal value of $\epsilon$ for OA, when applicable.
\item The power of the received useful signal for the O-terminal
  relative to the thermal noise, that is, the numerator of
  (\ref{eq:henormapr}).
\item The strength of the effective noises that affect performance of
  the O-terminals relative to the thermal noise, that is, the first
  two terms in the denominator of (\ref{eq:henormapr}).
  \end{itemize}
From the table, we can infer that depending on the operating scenario,
the main impairment is either thermal noise or interference from the
B-terminal transmission; sufficient pilots are allocated on the
downlink. Yet, the effects of channel estimation errors are not
negligible.

\begin{table*}[p]
 \begin{center}
{
\tabulinesep=1mm
\begin{tabu}{|c|c|c|c|c|c|c|c|c|}
  \hline
 & $\frac{\rho_o^\jbb}{\rho_b^\jbb}$  & $\rho_d^\jbb$  & $\rho_b^\jbb$  & $\rho_o^\jbb$  & optimal $\epsilon$  &  $ \frac{M'-1}{M'} 
    \cdot \rho_o\beta_o\cdot \frac{ \tau_p^o\rho_o\beta_o}{M'+\tau_p^o\rho_o\beta_o}$  [dB]  & $  \rho_o\beta_o \cdot \frac{M'}{M'+\tau_p^o\rho_o\beta_o}        $ [dB]    & $   \rho'_b\beta_o $ [dB] \\
& [dB]  &  [dB] &  [dB] &  [dB] &   & (power of useful signal  &  (channel estimation  & (B-terminal interference \\
&   &  &  &  &  & relative to  &   error relative to  & power relative to \\  
&   &  &  &  &  &  noise variance)  &   noise variance)  & noise variance) \\  
\hline
Figure~\ref{fig:a}, JBB$'$ & $11.0$ & $7.3$ & $-4.0 $   & $7.0 $  &  N/A  & $ 5.7$ & $ -2.1$  & $ -3.8$ \\ 
Figure~\ref{fig:a}, OA  & $12.5$ & $10.6$  & $-2.1$ & $ 10.3$ & $0.45$ & $  9.4 $ & $-1.8$ & $-\infty$ \\ 
Figure~\ref{fig:b}, JBB$'$  & $10.0$    &  $14.9$  &  $4.5$  & $14.5$ & N/A & $ 2.9$ &   $-2.5$ & $-5.3$ \\ 
Figure~\ref{fig:b}, OA  &  $10.7$   &  $17.3 $  & $ 6.3$ & $ 17.0 $ & $0.39$ & $ 5.7 $ & $ -2.1$ & $-\infty$  \\ 
Figure~\ref{fig:c}, JBB$'$   & $ 7.4$  & $ 10.8$& $ 2.7$ & $ 10.1 $ & N/A & $ 9.1 $ & $ -1.8 $ & $ 2.9$ \\ 
Figure~\ref{fig:c}, OA  & $ 9.6$  & $ 13.7$  & $ 3.7$ & $ 13.2$ & $0.41$ & $ 12.4 $ & $ -1.7$ & $-\infty$ \\ 
Figure~\ref{fig:e1}, JBB$'$   & $ 5.8$  & $ 0.9$& $ -6.0$ & $ -0.2$ & N/A & $ 9.3 $ & $ -1.8 $ & $ 4.6$ \\ 
Figure~\ref{fig:e1}, OA  & $ 8.3$  & $ 3.5$  & $ -5.4$ & $ 2.9$ & $0.37$ & $ 12.4 $ & $ -1.7$ & $-\infty$ \\
 \hline
\end{tabu}
}
\end{center}
\caption{The optimal value $\epsilon$ for OA, and the strength of the
  numerator and the first two terms in the denominator of
  (\ref{eq:henormapr}), for relevant operating points.}\label{table:1}
\end{table*}

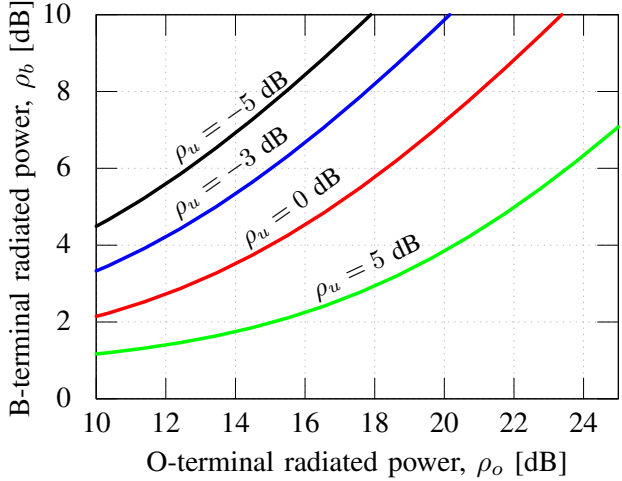
\begin{figure}[t!]
{\centerline{\scalebox{1.1}{\begin{tikzpicture}[gnuplot]
\path (0.000,0.000) rectangle (8.000,6.000);
\gpfill{rgb color={1.000,1.000,1.000}} (1.136,0.985)--(7.446,0.985)--(7.446,5.630)--(1.136,5.630)--cycle;
\gpcolor{color=gp lt color border}
\gpsetlinetype{gp lt border}
\gpsetlinewidth{1.00}
\draw[gp path] (1.136,0.985)--(1.136,5.630)--(7.446,5.630)--(7.446,0.985)--cycle;
\gpcolor{color=gp lt color axes}
\gpsetlinetype{gp lt axes}
\gpsetlinewidth{0.50}
\draw[gp path] (1.136,0.985)--(7.447,0.985);
\gpcolor{color=gp lt color border}
\gpsetlinetype{gp lt border}
\draw[gp path] (1.136,0.985)--(1.387,0.985);
\draw[gp path] (7.447,0.985)--(7.196,0.985);
\gpcolor{rgb color={0.000,0.000,0.000}}
\node[gp node right,font={\fontsize{10pt}{12pt}\selectfont}] at (0.952,0.985) {0};
\gpcolor{color=gp lt color axes}
\gpsetlinetype{gp lt axes}
\draw[gp path] (1.136,1.914)--(7.447,1.914);
\gpcolor{color=gp lt color border}
\gpsetlinetype{gp lt border}
\draw[gp path] (1.136,1.914)--(1.387,1.914);
\draw[gp path] (7.447,1.914)--(7.196,1.914);
\gpcolor{rgb color={0.000,0.000,0.000}}
\node[gp node right,font={\fontsize{10pt}{12pt}\selectfont}] at (0.952,1.914) {2};
\gpcolor{color=gp lt color axes}
\gpsetlinetype{gp lt axes}
\draw[gp path] (1.136,2.843)--(7.447,2.843);
\gpcolor{color=gp lt color border}
\gpsetlinetype{gp lt border}
\draw[gp path] (1.136,2.843)--(1.387,2.843);
\draw[gp path] (7.447,2.843)--(7.196,2.843);
\gpcolor{rgb color={0.000,0.000,0.000}}
\node[gp node right,font={\fontsize{10pt}{12pt}\selectfont}] at (0.952,2.843) {4};
\gpcolor{color=gp lt color axes}
\gpsetlinetype{gp lt axes}
\draw[gp path] (1.136,3.773)--(7.447,3.773);
\gpcolor{color=gp lt color border}
\gpsetlinetype{gp lt border}
\draw[gp path] (1.136,3.773)--(1.387,3.773);
\draw[gp path] (7.447,3.773)--(7.196,3.773);
\gpcolor{rgb color={0.000,0.000,0.000}}
\node[gp node right,font={\fontsize{10pt}{12pt}\selectfont}] at (0.952,3.773) {6};
\gpcolor{color=gp lt color axes}
\gpsetlinetype{gp lt axes}
\draw[gp path] (1.136,4.702)--(7.447,4.702);
\gpcolor{color=gp lt color border}
\gpsetlinetype{gp lt border}
\draw[gp path] (1.136,4.702)--(1.387,4.702);
\draw[gp path] (7.447,4.702)--(7.196,4.702);
\gpcolor{rgb color={0.000,0.000,0.000}}
\node[gp node right,font={\fontsize{10pt}{12pt}\selectfont}] at (0.952,4.702) {8};
\gpcolor{color=gp lt color axes}
\gpsetlinetype{gp lt axes}
\draw[gp path] (1.136,5.631)--(7.447,5.631);
\gpcolor{color=gp lt color border}
\gpsetlinetype{gp lt border}
\draw[gp path] (1.136,5.631)--(1.387,5.631);
\draw[gp path] (7.447,5.631)--(7.196,5.631);
\gpcolor{rgb color={0.000,0.000,0.000}}
\node[gp node right,font={\fontsize{10pt}{12pt}\selectfont}] at (0.952,5.631) {10};
\gpcolor{color=gp lt color axes}
\gpsetlinetype{gp lt axes}
\draw[gp path] (1.136,0.985)--(1.136,5.631);
\gpcolor{color=gp lt color border}
\gpsetlinetype{gp lt border}
\draw[gp path] (1.136,0.985)--(1.136,1.236);
\draw[gp path] (1.136,5.631)--(1.136,5.380);
\gpcolor{rgb color={0.000,0.000,0.000}}
\node[gp node center,font={\fontsize{10pt}{12pt}\selectfont}] at (1.136,0.677) {10};
\gpcolor{color=gp lt color axes}
\gpsetlinetype{gp lt axes}
\draw[gp path] (1.977,0.985)--(1.977,5.631);
\gpcolor{color=gp lt color border}
\gpsetlinetype{gp lt border}
\draw[gp path] (1.977,0.985)--(1.977,1.236);
\draw[gp path] (1.977,5.631)--(1.977,5.380);
\gpcolor{rgb color={0.000,0.000,0.000}}
\node[gp node center,font={\fontsize{10pt}{12pt}\selectfont}] at (1.977,0.677) {12};
\gpcolor{color=gp lt color axes}
\gpsetlinetype{gp lt axes}
\draw[gp path] (2.819,0.985)--(2.819,5.631);
\gpcolor{color=gp lt color border}
\gpsetlinetype{gp lt border}
\draw[gp path] (2.819,0.985)--(2.819,1.236);
\draw[gp path] (2.819,5.631)--(2.819,5.380);
\gpcolor{rgb color={0.000,0.000,0.000}}
\node[gp node center,font={\fontsize{10pt}{12pt}\selectfont}] at (2.819,0.677) {14};
\gpcolor{color=gp lt color axes}
\gpsetlinetype{gp lt axes}
\draw[gp path] (3.660,0.985)--(3.660,5.631);
\gpcolor{color=gp lt color border}
\gpsetlinetype{gp lt border}
\draw[gp path] (3.660,0.985)--(3.660,1.236);
\draw[gp path] (3.660,5.631)--(3.660,5.380);
\gpcolor{rgb color={0.000,0.000,0.000}}
\node[gp node center,font={\fontsize{10pt}{12pt}\selectfont}] at (3.660,0.677) {16};
\gpcolor{color=gp lt color axes}
\gpsetlinetype{gp lt axes}
\draw[gp path] (4.502,0.985)--(4.502,5.631);
\gpcolor{color=gp lt color border}
\gpsetlinetype{gp lt border}
\draw[gp path] (4.502,0.985)--(4.502,1.236);
\draw[gp path] (4.502,5.631)--(4.502,5.380);
\gpcolor{rgb color={0.000,0.000,0.000}}
\node[gp node center,font={\fontsize{10pt}{12pt}\selectfont}] at (4.502,0.677) {18};
\gpcolor{color=gp lt color axes}
\gpsetlinetype{gp lt axes}
\draw[gp path] (5.343,0.985)--(5.343,5.631);
\gpcolor{color=gp lt color border}
\gpsetlinetype{gp lt border}
\draw[gp path] (5.343,0.985)--(5.343,1.236);
\draw[gp path] (5.343,5.631)--(5.343,5.380);
\gpcolor{rgb color={0.000,0.000,0.000}}
\node[gp node center,font={\fontsize{10pt}{12pt}\selectfont}] at (5.343,0.677) {20};
\gpcolor{color=gp lt color axes}
\gpsetlinetype{gp lt axes}
\draw[gp path] (6.185,0.985)--(6.185,5.631);
\gpcolor{color=gp lt color border}
\gpsetlinetype{gp lt border}
\draw[gp path] (6.185,0.985)--(6.185,1.236);
\draw[gp path] (6.185,5.631)--(6.185,5.380);
\gpcolor{rgb color={0.000,0.000,0.000}}
\node[gp node center,font={\fontsize{10pt}{12pt}\selectfont}] at (6.185,0.677) {22};
\gpcolor{color=gp lt color axes}
\gpsetlinetype{gp lt axes}
\draw[gp path] (7.026,0.985)--(7.026,5.631);
\gpcolor{color=gp lt color border}
\gpsetlinetype{gp lt border}
\draw[gp path] (7.026,0.985)--(7.026,1.236);
\draw[gp path] (7.026,5.631)--(7.026,5.380);
\gpcolor{rgb color={0.000,0.000,0.000}}
\node[gp node center,font={\fontsize{10pt}{12pt}\selectfont}] at (7.026,0.677) {24};
\gpcolor{color=gp lt color border}
\draw[gp path] (1.136,5.631)--(1.136,0.985)--(7.447,0.985)--(7.447,5.631)--cycle;
\gpcolor{rgb color={0.000,0.000,0.000}}
\node[gp node center,rotate=90,font={\fontsize{10pt}{12pt}\selectfont}] at (0.246,3.308) {B-terminal radiated power, $\rho_b$ [dB]};
\node[gp node center,font={\fontsize{10pt}{12pt}\selectfont}] at (4.291,0.215) {O-terminal radiated power, $\rho_o$ [dB]};
\gpsetlinetype{gp lt plot 0}
\draw[gp path,very thick] (1.136,3.072)--(1.281,3.150)--(1.623,3.355)--(1.716,3.412)--(2.152,3.703)%
  --(2.326,3.829)--(2.587,4.021)--(2.948,4.304)--(3.022,4.362)--(3.457,4.726)--(3.516,4.778)%
  --(3.893,5.110)--(4.047,5.252)--(4.328,5.511)--(4.454,5.631);
\gpcolor{rgb color={0.000,0.000,1.000}}
\draw[gp path,very thick] (1.136,2.532)--(1.281,2.593)--(1.716,2.803)--(1.859,2.881)--(2.152,3.042)%
  --(2.587,3.309)--(2.655,3.355)--(3.022,3.605)--(3.327,3.829)--(3.457,3.926)--(3.893,4.271)%
  --(3.931,4.304)--(4.328,4.639)--(4.484,4.778)--(4.763,5.026)--(5.007,5.252)--(5.198,5.429)%
  --(5.409,5.631);
\gpcolor{rgb color={1.000,0.000,0.000}}
\draw[gp path,very thick] (1.136,1.982)--(1.281,2.020)--(1.716,2.157)--(2.152,2.318)--(2.358,2.407)%
  --(2.587,2.506)--(3.022,2.722)--(3.307,2.881)--(3.457,2.966)--(3.893,3.238)--(4.062,3.355)%
  --(4.328,3.538)--(4.716,3.829)--(4.763,3.864)--(5.198,4.214)--(5.303,4.304)--(5.633,4.585)%
  --(5.848,4.778)--(6.069,4.975)--(6.365,5.252)--(6.504,5.382)--(6.762,5.631);
\gpcolor{rgb color={0.000,1.000,0.000}}
\draw[gp path,very thick] (1.136,1.528)--(1.281,1.542)--(1.716,1.597)--(2.152,1.664)--(2.587,1.746)%
  --(3.022,1.845)--(3.340,1.933)--(3.457,1.965)--(3.893,2.107)--(4.328,2.275)--(4.625,2.407)%
  --(4.763,2.469)--(5.198,2.691)--(5.529,2.881)--(5.633,2.941)--(6.069,3.220)--(6.261,3.355)%
  --(6.504,3.526)--(6.902,3.829)--(6.939,3.858)--(7.374,4.213)--(7.447,4.275);
\gpcolor{rgb color={0.000,0.000,0.000}}
\node[gp node left,rotate=35,font={\fontsize{10pt}{12pt}\selectfont}] at (1.977,3.873) {\small $\rho_u=-5$~dB};
\node[gp node left,rotate=35,font={\fontsize{10pt}{12pt}\selectfont}] at (1.977,3.2) {\small $  \rho_u=-3$~dB};
\node[gp node left,rotate=35,font={\fontsize{10pt}{12pt}\selectfont}] at (2.819,2.843) {\small $  \rho_u=0$~dB};
\node[gp node left,rotate=25,font={\fontsize{10pt}{12pt}\selectfont}] at (3.660,2.25) {\small $\rho_u=5$~dB};
\gpdefrectangularnode{gp plot 1}{\pgfpoint{1.136cm}{0.985cm}}{\pgfpoint{7.447cm}{5.631cm}}
\end{tikzpicture}
}}}
\caption{\label{fig:e2} Example of required B-terminal power $\rho_b$
  for given O-terminal power $\rho_o$ in order to maintain a
  B-terminal sum-rate of $20$ b/s/Hz with $M=100$ antennas and $K=10$
  terminals, for different uplink pilot quality $\rho_u$. The channel
  coherence was $\tau_c=500$ symbols of which $\tau_p^u=30$ were spent
  on uplink pilots. }
\end{figure}

As an additional illustration, Figure~\ref{fig:e2} shows the required
B-terminal power $\rho_b$ for given O-terminal power $\rho_o$ in order
to maintain a B-terminal sum-rate of $20$ b/s/Hz with $M=100$ antennas
and $K=10$ terminals (that is, $2$ b/s/Hz/terminal). The channel
coherence was $\tau_c=500$ symbols of which $\tau_p^u=30$ were spent
on uplink pilots. Results are shown for different uplink pilot SNR
$\rho_u$. It can be seen that the better uplink pilot quality, the
more accurate channel state information is available to the
B-terminals and the less B-terminal power is required to maintain the
same rate. This is expected, because the larger $\rho_u$ is, the
closer is $\gamma_k$ to $\beta_k$ and the less is the leakage power in
(\ref{eq:leakagemag}).

\begin{figure}[t!]
{\centerline{\scalebox{1.1}{\begin{tikzpicture}[gnuplot]
\path (0.000,0.000) rectangle (8.000,6.000);
\gpfill{rgb color={1.000,1.000,1.000}} (1.136,0.985)--(7.446,0.985)--(7.446,5.630)--(1.136,5.630)--cycle;
\gpcolor{color=gp lt color border}
\gpsetlinetype{gp lt border}
\gpsetlinewidth{1.00}
\draw[gp path] (1.136,0.985)--(1.136,5.630)--(7.446,5.630)--(7.446,0.985)--cycle;
\gpcolor{color=gp lt color axes}
\gpsetlinetype{gp lt axes}
\gpsetlinewidth{0.50}
\draw[gp path] (1.136,0.985)--(7.447,0.985);
\gpcolor{color=gp lt color border}
\gpsetlinetype{gp lt border}
\draw[gp path] (1.136,0.985)--(1.387,0.985);
\draw[gp path] (7.447,0.985)--(7.196,0.985);
\gpcolor{rgb color={0.000,0.000,0.000}}
\node[gp node right,font={\fontsize{10pt}{12pt}\selectfont}] at (0.952,0.985) {0};
\gpcolor{color=gp lt color axes}
\gpsetlinetype{gp lt axes}
\draw[gp path] (1.136,2.147)--(7.447,2.147);
\gpcolor{color=gp lt color border}
\gpsetlinetype{gp lt border}
\draw[gp path] (1.136,2.147)--(1.387,2.147);
\draw[gp path] (7.447,2.147)--(7.196,2.147);
\gpcolor{rgb color={0.000,0.000,0.000}}
\node[gp node right,font={\fontsize{10pt}{12pt}\selectfont}] at (0.952,2.147) {5};
\gpcolor{color=gp lt color axes}
\gpsetlinetype{gp lt axes}
\draw[gp path] (1.136,3.308)--(7.447,3.308);
\gpcolor{color=gp lt color border}
\gpsetlinetype{gp lt border}
\draw[gp path] (1.136,3.308)--(1.387,3.308);
\draw[gp path] (7.447,3.308)--(7.196,3.308);
\gpcolor{rgb color={0.000,0.000,0.000}}
\node[gp node right,font={\fontsize{10pt}{12pt}\selectfont}] at (0.952,3.308) {10};
\gpcolor{color=gp lt color axes}
\gpsetlinetype{gp lt axes}
\draw[gp path] (1.136,4.470)--(7.447,4.470);
\gpcolor{color=gp lt color border}
\gpsetlinetype{gp lt border}
\draw[gp path] (1.136,4.470)--(1.387,4.470);
\draw[gp path] (7.447,4.470)--(7.196,4.470);
\gpcolor{rgb color={0.000,0.000,0.000}}
\node[gp node right,font={\fontsize{10pt}{12pt}\selectfont}] at (0.952,4.470) {15};
\gpcolor{color=gp lt color axes}
\gpsetlinetype{gp lt axes}
\draw[gp path] (1.136,5.631)--(7.447,5.631);
\gpcolor{color=gp lt color border}
\gpsetlinetype{gp lt border}
\draw[gp path] (1.136,5.631)--(1.387,5.631);
\draw[gp path] (7.447,5.631)--(7.196,5.631);
\gpcolor{rgb color={0.000,0.000,0.000}}
\node[gp node right,font={\fontsize{10pt}{12pt}\selectfont}] at (0.952,5.631) {20};
\gpcolor{color=gp lt color axes}
\gpsetlinetype{gp lt axes}
\draw[gp path] (1.136,0.985)--(1.136,5.631);
\gpcolor{color=gp lt color border}
\gpsetlinetype{gp lt border}
\draw[gp path] (1.136,0.985)--(1.136,1.236);
\draw[gp path] (1.136,5.631)--(1.136,5.380);
\gpcolor{rgb color={0.000,0.000,0.000}}
\node[gp node center,font={\fontsize{10pt}{12pt}\selectfont}] at (1.136,0.677) {0};
\gpcolor{color=gp lt color axes}
\gpsetlinetype{gp lt axes}
\draw[gp path] (2.714,0.985)--(2.714,5.631);
\gpcolor{color=gp lt color border}
\gpsetlinetype{gp lt border}
\draw[gp path] (2.714,0.985)--(2.714,1.236);
\draw[gp path] (2.714,5.631)--(2.714,5.380);
\gpcolor{rgb color={0.000,0.000,0.000}}
\node[gp node center,font={\fontsize{10pt}{12pt}\selectfont}] at (2.714,0.677) {5};
\gpcolor{color=gp lt color axes}
\gpsetlinetype{gp lt axes}
\draw[gp path] (4.292,0.985)--(4.292,5.631);
\gpcolor{color=gp lt color border}
\gpsetlinetype{gp lt border}
\draw[gp path] (4.292,0.985)--(4.292,1.236);
\draw[gp path] (4.292,5.631)--(4.292,5.380);
\gpcolor{rgb color={0.000,0.000,0.000}}
\node[gp node center,font={\fontsize{10pt}{12pt}\selectfont}] at (4.292,0.677) {10};
\gpcolor{color=gp lt color axes}
\gpsetlinetype{gp lt axes}
\draw[gp path] (5.869,0.985)--(5.869,5.631);
\gpcolor{color=gp lt color border}
\gpsetlinetype{gp lt border}
\draw[gp path] (5.869,0.985)--(5.869,1.236);
\draw[gp path] (5.869,5.631)--(5.869,5.380);
\gpcolor{rgb color={0.000,0.000,0.000}}
\node[gp node center,font={\fontsize{10pt}{12pt}\selectfont}] at (5.869,0.677) {15};
\gpcolor{color=gp lt color axes}
\gpsetlinetype{gp lt axes}
\draw[gp path] (7.447,0.985)--(7.447,5.631);
\gpcolor{color=gp lt color border}
\gpsetlinetype{gp lt border}
\draw[gp path] (7.447,0.985)--(7.447,1.236);
\draw[gp path] (7.447,5.631)--(7.447,5.380);
\gpcolor{rgb color={0.000,0.000,0.000}}
\node[gp node center,font={\fontsize{10pt}{12pt}\selectfont}] at (7.447,0.677) {20};
\gpcolor{color=gp lt color border}
\draw[gp path] (1.136,5.631)--(1.136,0.985)--(7.447,0.985)--(7.447,5.631)--cycle;
\gpcolor{rgb color={0.000,0.000,0.000}}
\node[gp node center,rotate=90,font={\fontsize{10pt}{12pt}\selectfont}] at (0.246,3.308) {total radiated power, $\rho_d^\jbb$ [dB]};
\node[gp node center,font={\fontsize{10pt}{12pt}\selectfont}] at (4.291,0.215) {power ratio $\rho_o^\jbb/\rho_b^\jbb$ [dB]};
\gpsetlinetype{gp lt plot 0}
\draw[gp path,very thick,postaction={decorate,decoration={text along path,text align={left indent=2mm}, text=|\scriptsize|B-term. 2 b/s/Hz,text align=left,raise=3pt}}] (2.671,0.985)--(2.797,1.067)--(3.032,1.230)--(3.129,1.300)--(3.359,1.474)%
  --(3.461,1.555)--(3.659,1.719)--(3.793,1.837)--(3.933,1.963)--(4.125,2.151)--(4.182,2.208)%
  --(4.406,2.452)--(4.458,2.515)--(4.606,2.697)--(4.782,2.941)--(4.790,2.953)--(4.936,3.186)%
  --(5.068,3.430)--(5.122,3.547)--(5.180,3.675)--(5.275,3.919)--(5.353,4.164)--(5.418,4.408)%
  --(5.454,4.577)--(5.470,4.653)--(5.513,4.897)--(5.547,5.142)--(5.575,5.386)--(5.597,5.631);
\gpcolor{rgb color={1.000,0.000,0.000}}
\gpsetlinetype{gp lt plot 1}
\draw[gp path,very thick] (2.516,5.631)--(2.541,5.386)--(2.573,5.142)--(2.614,4.897)--(2.668,4.653)%
  --(2.739,4.408)--(2.797,4.253)--(2.831,4.164)--(2.951,3.919)--(3.112,3.675)--(3.129,3.654)%
  --(3.335,3.430)--(3.461,3.323)--(3.654,3.186)--(3.793,3.104)--(4.125,2.950)--(4.149,2.941)%
  --(4.458,2.838)--(4.790,2.755)--(5.097,2.697)--(5.122,2.692)--(5.454,2.643)--(5.786,2.606)%
  --(6.118,2.577)--(6.451,2.555)--(6.783,2.537)--(7.115,2.524)--(7.447,2.514);
\gpcolor{rgb color={0.000,0.000,1.000}}
\draw[gp path,very thick,postaction={decorate,decoration={text along path, text align={left indent=5mm},text=|\scriptsize|O-term. 0.75 b/s/Hz OA,text align=left,raise=3pt}}] (2.743,5.631)--(2.797,5.550)--(2.909,5.386)--(3.095,5.142)--(3.129,5.100)%
  --(3.300,4.897)--(3.461,4.723)--(3.529,4.653)--(3.786,4.408)--(3.793,4.402)--(4.080,4.164)%
  --(4.125,4.128)--(4.409,3.919)--(4.458,3.885)--(4.790,3.680)--(4.798,3.675)--(5.122,3.500)%
  --(5.266,3.430)--(5.454,3.343)--(5.786,3.208)--(5.849,3.186)--(6.118,3.091)--(6.451,2.991)%
  --(6.639,2.941)--(6.783,2.904)--(7.115,2.830)--(7.447,2.767);
\gpcolor{rgb color={1.000,0.000,0.000}}
\gpsetlinetype{gp lt plot 0}
\draw[gp path,very thick,postaction={decorate,decoration={text along path, text align={left indent=10mm},text=|\scriptsize|O-term. 0.75 b/s/Hz JBB$'$\ ,text align=left,raise=-8pt}}] (2.359,5.631)--(2.410,5.386)--(2.410,5.142)--(2.440,4.897)--(2.465,4.811)%
  --(2.511,4.653)--(2.571,4.408)--(2.662,4.164)--(2.778,3.919)--(2.797,3.896)--(2.949,3.675)%
  --(3.094,3.430)--(3.129,3.392)--(3.389,3.186)--(3.461,3.136)--(3.793,2.950)--(3.814,2.941)%
  --(4.125,2.831)--(4.458,2.718)--(4.545,2.697)--(4.790,2.636)--(5.122,2.573)--(5.454,2.538)%
  --(5.786,2.492)--(6.118,2.455)--(6.160,2.452)--(6.451,2.435)--(6.783,2.423)--(7.115,2.431)%
  --(7.447,2.427);
\gpcolor{rgb color={0.000,0.000,1.000}}
\draw[gp path,very thick] (2.704,5.631)--(2.797,5.487)--(2.860,5.386)--(3.039,5.142)--(3.129,5.030)%
  --(3.239,4.897)--(3.457,4.653)--(3.461,4.648)--(3.698,4.408)--(3.793,4.323)--(3.978,4.164)%
  --(4.125,4.046)--(4.296,3.919)--(4.458,3.810)--(4.657,3.675)--(4.790,3.590)--(5.096,3.430)%
  --(5.122,3.417)--(5.454,3.237)--(5.579,3.186)--(5.786,3.112)--(6.118,2.992)--(6.293,2.941)%
  --(6.451,2.894)--(6.783,2.795)--(7.115,2.725)--(7.301,2.697)--(7.447,2.672);
\gpdefrectangularnode{gp plot 1}{\pgfpoint{1.136cm}{0.985cm}}{\pgfpoint{7.447cm}{5.631cm}}
\end{tikzpicture}
}}}
\caption{\label{fig:a} Feasible operating points in terms of the total
  downlink power $\rho_d^\jbb=\rho_b^\jbb+ \rho_o^\jbb$ and the power
  ratio $\rho_o^\jbb/\rho_b^\jbb$, which yield a predetermined net
  rate for a B- and an O-terminal, both located at the cell border,
  for the case of {${M=100}$} antennas, $K=1$ B-terminal, $\tau_c=500$
  symbols channel coherence, $\rho_u=-3$ dB uplink SNR, $\tau_p^u=10$
  uplink pilot symbols, dimensionality $M'=7$ of the reduced channel,
  and $\tau_p^o=10$ downlink pilots.  The black line represents the
  expression (\ref{eq:b51}) for B-terminal rate with JBB$'$.  The
  solid red line represents the O-terminal rate with JBB$'$,
  (\ref{eq:orate1}) weighted to account for the pilot cost. The solid blue line represents
  the O-terminal rate with OA, (\ref{eq:Roortho}) weighted to account
  for the pilot cost  and
  optimized with respect to $\epsilon$.  The red and blue dashed lines represent the
  corresponding closed-form bounds (\ref{eq:henormapr}) respectively
  (\ref{eq:Roortho_appr}) on the O-terminal rates.  }
\end{figure}
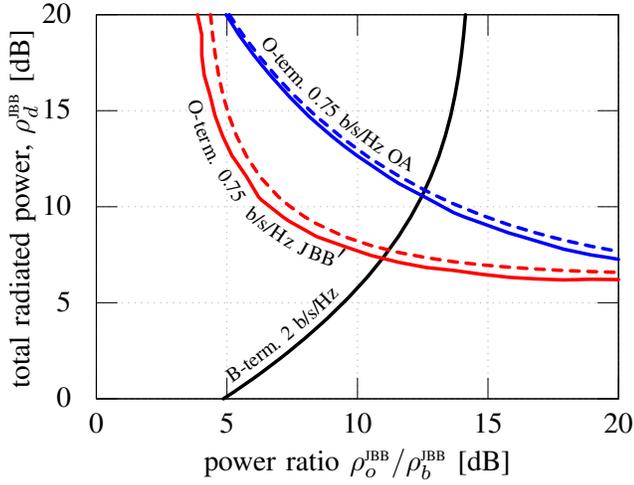
 
\begin{figure}[t!]
{\centerline{\scalebox{1.1}{\begin{tikzpicture}[gnuplot]
\path (0.000,0.000) rectangle (8.000,6.000);
\gpfill{rgb color={1.000,1.000,1.000}} (1.136,0.985)--(7.446,0.985)--(7.446,5.630)--(1.136,5.630)--cycle;
\gpcolor{color=gp lt color border}
\gpsetlinetype{gp lt border}
\gpsetlinewidth{1.00}
\draw[gp path] (1.136,0.985)--(1.136,5.630)--(7.446,5.630)--(7.446,0.985)--cycle;
\gpcolor{color=gp lt color axes}
\gpsetlinetype{gp lt axes}
\gpsetlinewidth{0.50}
\draw[gp path] (1.136,0.985)--(7.447,0.985);
\gpcolor{color=gp lt color border}
\gpsetlinetype{gp lt border}
\draw[gp path] (1.136,0.985)--(1.387,0.985);
\draw[gp path] (7.447,0.985)--(7.196,0.985);
\gpcolor{rgb color={0.000,0.000,0.000}}
\node[gp node right,font={\fontsize{10pt}{12pt}\selectfont}] at (0.952,0.985) {10};
\gpcolor{color=gp lt color axes}
\gpsetlinetype{gp lt axes}
\draw[gp path] (1.136,2.147)--(7.447,2.147);
\gpcolor{color=gp lt color border}
\gpsetlinetype{gp lt border}
\draw[gp path] (1.136,2.147)--(1.387,2.147);
\draw[gp path] (7.447,2.147)--(7.196,2.147);
\gpcolor{rgb color={0.000,0.000,0.000}}
\node[gp node right,font={\fontsize{10pt}{12pt}\selectfont}] at (0.952,2.147) {15};
\gpcolor{color=gp lt color axes}
\gpsetlinetype{gp lt axes}
\draw[gp path] (1.136,3.308)--(7.447,3.308);
\gpcolor{color=gp lt color border}
\gpsetlinetype{gp lt border}
\draw[gp path] (1.136,3.308)--(1.387,3.308);
\draw[gp path] (7.447,3.308)--(7.196,3.308);
\gpcolor{rgb color={0.000,0.000,0.000}}
\node[gp node right,font={\fontsize{10pt}{12pt}\selectfont}] at (0.952,3.308) {20};
\gpcolor{color=gp lt color axes}
\gpsetlinetype{gp lt axes}
\draw[gp path] (1.136,4.470)--(7.447,4.470);
\gpcolor{color=gp lt color border}
\gpsetlinetype{gp lt border}
\draw[gp path] (1.136,4.470)--(1.387,4.470);
\draw[gp path] (7.447,4.470)--(7.196,4.470);
\gpcolor{rgb color={0.000,0.000,0.000}}
\node[gp node right,font={\fontsize{10pt}{12pt}\selectfont}] at (0.952,4.470) {25};
\gpcolor{color=gp lt color axes}
\gpsetlinetype{gp lt axes}
\draw[gp path] (1.136,5.631)--(7.447,5.631);
\gpcolor{color=gp lt color border}
\gpsetlinetype{gp lt border}
\draw[gp path] (1.136,5.631)--(1.387,5.631);
\draw[gp path] (7.447,5.631)--(7.196,5.631);
\gpcolor{rgb color={0.000,0.000,0.000}}
\node[gp node right,font={\fontsize{10pt}{12pt}\selectfont}] at (0.952,5.631) {30};
\gpcolor{color=gp lt color axes}
\gpsetlinetype{gp lt axes}
\draw[gp path] (1.136,0.985)--(1.136,5.631);
\gpcolor{color=gp lt color border}
\gpsetlinetype{gp lt border}
\draw[gp path] (1.136,0.985)--(1.136,1.236);
\draw[gp path] (1.136,5.631)--(1.136,5.380);
\gpcolor{rgb color={0.000,0.000,0.000}}
\node[gp node center,font={\fontsize{10pt}{12pt}\selectfont}] at (1.136,0.677) {0};
\gpcolor{color=gp lt color axes}
\gpsetlinetype{gp lt axes}
\draw[gp path] (2.714,0.985)--(2.714,5.631);
\gpcolor{color=gp lt color border}
\gpsetlinetype{gp lt border}
\draw[gp path] (2.714,0.985)--(2.714,1.236);
\draw[gp path] (2.714,5.631)--(2.714,5.380);
\gpcolor{rgb color={0.000,0.000,0.000}}
\node[gp node center,font={\fontsize{10pt}{12pt}\selectfont}] at (2.714,0.677) {5};
\gpcolor{color=gp lt color axes}
\gpsetlinetype{gp lt axes}
\draw[gp path] (4.292,0.985)--(4.292,5.631);
\gpcolor{color=gp lt color border}
\gpsetlinetype{gp lt border}
\draw[gp path] (4.292,0.985)--(4.292,1.236);
\draw[gp path] (4.292,5.631)--(4.292,5.380);
\gpcolor{rgb color={0.000,0.000,0.000}}
\node[gp node center,font={\fontsize{10pt}{12pt}\selectfont}] at (4.292,0.677) {10};
\gpcolor{color=gp lt color axes}
\gpsetlinetype{gp lt axes}
\draw[gp path] (5.869,0.985)--(5.869,5.631);
\gpcolor{color=gp lt color border}
\gpsetlinetype{gp lt border}
\draw[gp path] (5.869,0.985)--(5.869,1.236);
\draw[gp path] (5.869,5.631)--(5.869,5.380);
\gpcolor{rgb color={0.000,0.000,0.000}}
\node[gp node center,font={\fontsize{10pt}{12pt}\selectfont}] at (5.869,0.677) {15};
\gpcolor{color=gp lt color axes}
\gpsetlinetype{gp lt axes}
\draw[gp path] (7.447,0.985)--(7.447,5.631);
\gpcolor{color=gp lt color border}
\gpsetlinetype{gp lt border}
\draw[gp path] (7.447,0.985)--(7.447,1.236);
\draw[gp path] (7.447,5.631)--(7.447,5.380);
\gpcolor{rgb color={0.000,0.000,0.000}}
\node[gp node center,font={\fontsize{10pt}{12pt}\selectfont}] at (7.447,0.677) {20};
\gpcolor{color=gp lt color border}
\draw[gp path] (1.136,5.631)--(1.136,0.985)--(7.447,0.985)--(7.447,5.631)--cycle;
\gpcolor{rgb color={0.000,0.000,0.000}}
\node[gp node center,rotate=90,font={\fontsize{10pt}{12pt}\selectfont}] at (0.246,3.308) {total radiated power, $\rho_d^\jbb$ [dB]};
\node[gp node center,font={\fontsize{10pt}{12pt}\selectfont}] at (4.291,0.215) {power ratio $\rho_o^\jbb/\rho_b^\jbb$ [dB]};
\gpsetlinetype{gp lt plot 0}
\draw[gp path,very thick,postaction={decorate,decoration={text along path,text align={left indent=27mm}, text=|\scriptsize|B-term. 20 b/s/Hz,text align=left,raise=3pt}}] (3.449,0.985)--(3.461,0.997)--(3.675,1.230)--(3.793,1.374)--(3.874,1.474)%
  --(4.046,1.719)--(4.125,1.847)--(4.195,1.963)--(4.322,2.208)--(4.428,2.452)--(4.458,2.532)%
  --(4.517,2.697)--(4.590,2.941)--(4.650,3.186)--(4.698,3.430)--(4.738,3.675)--(4.769,3.919)%
  --(4.790,4.119)--(4.794,4.164)--(4.814,4.408)--(4.830,4.653)--(4.843,4.897)--(4.853,5.142)%
  --(4.860,5.386)--(4.867,5.631);
\gpcolor{rgb color={1.000,0.000,0.000}}
\gpsetlinetype{gp lt plot 1}
\draw[gp path,very thick] (1.736,5.631)--(1.756,5.386)--(1.781,5.142)--(1.800,4.993)--(1.812,4.897)%
  --(1.850,4.653)--(1.899,4.408)--(1.962,4.164)--(2.045,3.919)--(2.132,3.719)--(2.152,3.675)%
  --(2.288,3.430)--(2.465,3.191)--(2.469,3.186)--(2.711,2.941)--(2.797,2.871)--(3.050,2.697)%
  --(3.129,2.651)--(3.461,2.491)--(3.563,2.452)--(3.793,2.372)--(4.125,2.281)--(4.458,2.213)%
  --(4.491,2.208)--(4.790,2.159)--(5.122,2.117)--(5.454,2.084)--(5.786,2.059)--(6.118,2.040)%
  --(6.451,2.024)--(6.783,2.013)--(7.115,2.004)--(7.447,1.996);
\gpcolor{rgb color={0.000,0.000,1.000}}
\draw[gp path,very thick] (1.555,5.631)--(1.681,5.386)--(1.800,5.173)--(1.816,5.142)--(1.951,4.897)%
  --(2.114,4.653)--(2.132,4.628)--(2.301,4.408)--(2.465,4.207)--(2.500,4.164)--(2.727,3.919)%
  --(2.797,3.852)--(2.992,3.675)--(3.129,3.561)--(3.302,3.430)--(3.461,3.318)--(3.671,3.186)%
  --(3.793,3.113)--(4.122,2.941)--(4.125,2.939)--(4.458,2.790)--(4.700,2.697)--(4.790,2.663)%
  --(5.122,2.554)--(5.454,2.463)--(5.498,2.452)--(5.786,2.382)--(6.118,2.314)--(6.451,2.256)%
  --(6.783,2.208)--(6.789,2.208)--(7.115,2.165)--(7.447,2.129);
\gpcolor{rgb color={1.000,0.000,0.000}}
\gpsetlinetype{gp lt plot 0}
\draw[gp path,very thick,postaction={decorate,decoration={text along path, text align={left indent=20mm},text=|\scriptsize|O-term. 0.5 b/s/Hz JBB$'$\ ,raise=-8pt}}] (1.579,5.631)--(1.579,5.386)--(1.610,5.142)--(1.644,4.897)--(1.676,4.653)%
  --(1.751,4.408)--(1.768,4.164)--(1.800,4.068)--(1.850,3.919)--(1.990,3.675)--(2.090,3.430)%
  --(2.132,3.362)--(2.260,3.186)--(2.442,2.941)--(2.465,2.922)--(2.790,2.697)--(2.797,2.692)%
  --(3.129,2.503)--(3.254,2.452)--(3.461,2.364)--(3.793,2.252)--(3.956,2.208)--(4.125,2.161)%
  --(4.458,2.092)--(4.790,2.039)--(5.122,1.998)--(5.454,1.986)--(5.615,1.963)--(5.786,1.937)%
  --(6.118,1.952)--(6.451,1.920)--(6.783,1.893)--(7.115,1.904)--(7.447,1.886);
\gpcolor{rgb color={0.000,0.000,1.000}}
\draw[gp path,very thick,postaction={decorate,decoration={text along path,text align={left indent=15mm},text=|\scriptsize|O-term. 0.5 b/s/Hz OA,text align=left,raise=5pt}}] (1.526,5.631)--(1.648,5.386)--(1.779,5.142)--(1.800,5.103)--(1.904,4.897)%
  --(2.070,4.653)--(2.132,4.564)--(2.245,4.408)--(2.428,4.164)--(2.465,4.124)--(2.675,3.919)%
  --(2.797,3.790)--(2.910,3.675)--(3.129,3.480)--(3.193,3.430)--(3.461,3.232)--(3.528,3.186)%
  --(3.793,3.016)--(3.941,2.941)--(4.125,2.857)--(4.458,2.711)--(4.492,2.697)--(4.790,2.571)%
  --(5.122,2.464)--(5.160,2.452)--(5.454,2.362)--(5.786,2.275)--(6.118,2.212)--(6.146,2.208)%
  --(6.451,2.153)--(6.783,2.090)--(7.115,2.056)--(7.447,2.014);
\gpdefrectangularnode{gp plot 1}{\pgfpoint{1.136cm}{0.985cm}}{\pgfpoint{7.447cm}{5.631cm}}
\end{tikzpicture}
}}}
\caption{\label{fig:b} Same as Figure~\ref{fig:a} but for $M=100$
  antennas serving $K=10$ B-terminals located at random in the cell.
  Here $\tau_p^u=30$, $\tau_p^o=10$ and $M'=7$. The O-terminals were
  located on the cell border, with an additional shadow margin of
  10~dB.}
\end{figure}
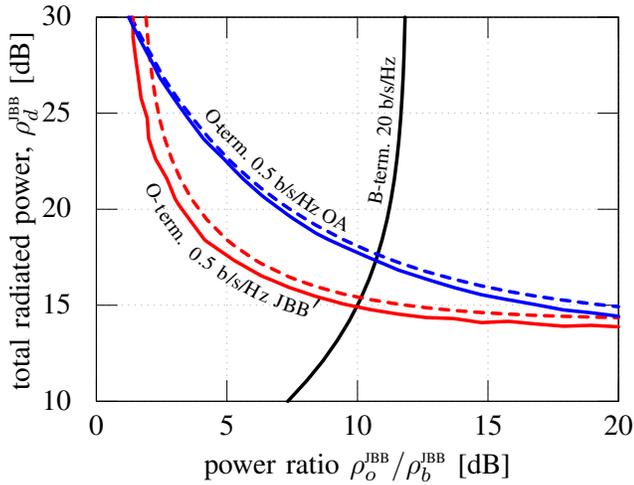

\begin{figure}[t!]
{\centerline{\scalebox{1.1}{\begin{tikzpicture}[gnuplot]
\path (0.000,0.000) rectangle (8.000,6.000);
\gpfill{rgb color={1.000,1.000,1.000}} (1.136,0.985)--(7.446,0.985)--(7.446,5.630)--(1.136,5.630)--cycle;
\gpcolor{color=gp lt color border}
\gpsetlinetype{gp lt border}
\gpsetlinewidth{1.00}
\draw[gp path] (1.136,0.985)--(1.136,5.630)--(7.446,5.630)--(7.446,0.985)--cycle;
\gpcolor{color=gp lt color axes}
\gpsetlinetype{gp lt axes}
\gpsetlinewidth{0.50}
\draw[gp path] (1.136,0.985)--(7.447,0.985);
\gpcolor{color=gp lt color border}
\gpsetlinetype{gp lt border}
\draw[gp path] (1.136,0.985)--(1.387,0.985);
\draw[gp path] (7.447,0.985)--(7.196,0.985);
\gpcolor{rgb color={0.000,0.000,0.000}}
\node[gp node right,font={\fontsize{10pt}{12pt}\selectfont}] at (0.952,0.985) {0};
\gpcolor{color=gp lt color axes}
\gpsetlinetype{gp lt axes}
\draw[gp path] (1.136,2.147)--(7.447,2.147);
\gpcolor{color=gp lt color border}
\gpsetlinetype{gp lt border}
\draw[gp path] (1.136,2.147)--(1.387,2.147);
\draw[gp path] (7.447,2.147)--(7.196,2.147);
\gpcolor{rgb color={0.000,0.000,0.000}}
\node[gp node right,font={\fontsize{10pt}{12pt}\selectfont}] at (0.952,2.147) {5};
\gpcolor{color=gp lt color axes}
\gpsetlinetype{gp lt axes}
\draw[gp path] (1.136,3.308)--(7.447,3.308);
\gpcolor{color=gp lt color border}
\gpsetlinetype{gp lt border}
\draw[gp path] (1.136,3.308)--(1.387,3.308);
\draw[gp path] (7.447,3.308)--(7.196,3.308);
\gpcolor{rgb color={0.000,0.000,0.000}}
\node[gp node right,font={\fontsize{10pt}{12pt}\selectfont}] at (0.952,3.308) {10};
\gpcolor{color=gp lt color axes}
\gpsetlinetype{gp lt axes}
\draw[gp path] (1.136,4.470)--(7.447,4.470);
\gpcolor{color=gp lt color border}
\gpsetlinetype{gp lt border}
\draw[gp path] (1.136,4.470)--(1.387,4.470);
\draw[gp path] (7.447,4.470)--(7.196,4.470);
\gpcolor{rgb color={0.000,0.000,0.000}}
\node[gp node right,font={\fontsize{10pt}{12pt}\selectfont}] at (0.952,4.470) {15};
\gpcolor{color=gp lt color axes}
\gpsetlinetype{gp lt axes}
\draw[gp path] (1.136,5.631)--(7.447,5.631);
\gpcolor{color=gp lt color border}
\gpsetlinetype{gp lt border}
\draw[gp path] (1.136,5.631)--(1.387,5.631);
\draw[gp path] (7.447,5.631)--(7.196,5.631);
\gpcolor{rgb color={0.000,0.000,0.000}}
\node[gp node right,font={\fontsize{10pt}{12pt}\selectfont}] at (0.952,5.631) {20};
\gpcolor{color=gp lt color axes}
\gpsetlinetype{gp lt axes}
\draw[gp path] (1.136,0.985)--(1.136,5.631);
\gpcolor{color=gp lt color border}
\gpsetlinetype{gp lt border}
\draw[gp path] (1.136,0.985)--(1.136,1.236);
\draw[gp path] (1.136,5.631)--(1.136,5.380);
\gpcolor{rgb color={0.000,0.000,0.000}}
\node[gp node center,font={\fontsize{10pt}{12pt}\selectfont}] at (1.136,0.677) {0};
\gpcolor{color=gp lt color axes}
\gpsetlinetype{gp lt axes}
\draw[gp path] (2.714,0.985)--(2.714,5.631);
\gpcolor{color=gp lt color border}
\gpsetlinetype{gp lt border}
\draw[gp path] (2.714,0.985)--(2.714,1.236);
\draw[gp path] (2.714,5.631)--(2.714,5.380);
\gpcolor{rgb color={0.000,0.000,0.000}}
\node[gp node center,font={\fontsize{10pt}{12pt}\selectfont}] at (2.714,0.677) {5};
\gpcolor{color=gp lt color axes}
\gpsetlinetype{gp lt axes}
\draw[gp path] (4.292,0.985)--(4.292,5.631);
\gpcolor{color=gp lt color border}
\gpsetlinetype{gp lt border}
\draw[gp path] (4.292,0.985)--(4.292,1.236);
\draw[gp path] (4.292,5.631)--(4.292,5.380);
\gpcolor{rgb color={0.000,0.000,0.000}}
\node[gp node center,font={\fontsize{10pt}{12pt}\selectfont}] at (4.292,0.677) {10};
\gpcolor{color=gp lt color axes}
\gpsetlinetype{gp lt axes}
\draw[gp path] (5.869,0.985)--(5.869,5.631);
\gpcolor{color=gp lt color border}
\gpsetlinetype{gp lt border}
\draw[gp path] (5.869,0.985)--(5.869,1.236);
\draw[gp path] (5.869,5.631)--(5.869,5.380);
\gpcolor{rgb color={0.000,0.000,0.000}}
\node[gp node center,font={\fontsize{10pt}{12pt}\selectfont}] at (5.869,0.677) {15};
\gpcolor{color=gp lt color axes}
\gpsetlinetype{gp lt axes}
\draw[gp path] (7.447,0.985)--(7.447,5.631);
\gpcolor{color=gp lt color border}
\gpsetlinetype{gp lt border}
\draw[gp path] (7.447,0.985)--(7.447,1.236);
\draw[gp path] (7.447,5.631)--(7.447,5.380);
\gpcolor{rgb color={0.000,0.000,0.000}}
\node[gp node center,font={\fontsize{10pt}{12pt}\selectfont}] at (7.447,0.677) {20};
\gpcolor{color=gp lt color border}
\draw[gp path] (1.136,5.631)--(1.136,0.985)--(7.447,0.985)--(7.447,5.631)--cycle;
\gpcolor{rgb color={0.000,0.000,0.000}}
\node[gp node center,rotate=90,font={\fontsize{10pt}{12pt}\selectfont}] at (0.246,3.308) {total radiated power, $\rho_d^\jbb$ [dB]};
\node[gp node center,font={\fontsize{10pt}{12pt}\selectfont}] at (4.291,0.215) {power ratio $\rho_o^\jbb/\rho_b^\jbb$ [dB]};
\gpsetlinetype{gp lt plot 0}
\draw[gp path,very thick,postaction={decorate,decoration={text along path,reverse path,text align={left indent=5mm}, text=|\scriptsize|B-term. 50 b/s/Hz,text align=left,raise=3pt}}] (4.997,5.631)--(4.906,5.386)--(4.798,5.142)--(4.790,5.127)--(4.669,4.897)%
  --(4.520,4.653)--(4.458,4.564)--(4.347,4.408)--(4.150,4.164)--(4.125,4.137)--(3.927,3.919)%
  --(3.793,3.786)--(3.679,3.675)--(3.461,3.480)--(3.404,3.430)--(3.129,3.208)--(3.100,3.186)%
  --(2.797,2.964)--(2.764,2.941)--(2.465,2.743)--(2.390,2.697)--(2.132,2.543)--(1.967,2.452)%
  --(1.800,2.363)--(1.482,2.208)--(1.468,2.201)--(1.136,2.058);
\gpcolor{rgb color={1.000,0.000,0.000}}
\gpsetlinetype{gp lt plot 1}
\draw[gp path,very thick] (2.803,5.631)--(2.831,5.386)--(2.868,5.142)--(2.915,4.897)--(2.977,4.653)%
  --(3.058,4.408)--(3.129,4.243)--(3.165,4.164)--(3.308,3.919)--(3.461,3.723)--(3.504,3.675)%
  --(3.782,3.430)--(3.793,3.422)--(4.125,3.225)--(4.211,3.186)--(4.458,3.086)--(4.790,2.984)%
  --(4.971,2.941)--(5.122,2.908)--(5.454,2.850)--(5.786,2.806)--(6.118,2.773)--(6.451,2.747)%
  --(6.783,2.727)--(7.115,2.711)--(7.447,2.699);
\gpcolor{rgb color={0.000,0.000,1.000}}
\draw[gp path,very thick,postaction={decorate,decoration={text along path, text align={left indent=25mm},text=|\scriptsize|O-term. 0.75 b/s/Hz OA,text align=left,raise=3pt}}] (3.057,5.631)--(3.129,5.525)--(3.224,5.386)--(3.399,5.142)--(3.461,5.058)%
  --(3.587,4.897)--(3.790,4.653)--(3.793,4.650)--(4.007,4.408)--(4.125,4.286)--(4.252,4.164)%
  --(4.458,3.978)--(4.528,3.919)--(4.790,3.717)--(4.851,3.675)--(5.122,3.503)--(5.254,3.430)%
  --(5.454,3.329)--(5.786,3.189)--(5.796,3.186)--(6.118,3.077)--(6.451,2.988)--(6.668,2.941)%
  --(6.783,2.918)--(7.115,2.861)--(7.447,2.817);
\gpcolor{rgb color={1.000,0.000,0.000}}
\gpsetlinetype{gp lt plot 0}
\draw[gp path,very thick,postaction={decorate,decoration={text along path, text align={left indent=25mm},text=|\scriptsize|O-term. 0.75 b/s/Hz JBB$'$\ ,text align=left,raise=-8pt}}] (2.648,5.631)--(2.667,5.386)--(2.730,5.142)--(2.765,4.897)--(2.797,4.781)%
  --(2.825,4.653)--(2.886,4.408)--(2.967,4.164)--(3.086,3.919)--(3.129,3.844)--(3.274,3.675)%
  --(3.461,3.519)--(3.555,3.430)--(3.793,3.258)--(3.904,3.186)--(4.125,3.062)--(4.458,2.958)%
  --(4.508,2.941)--(4.790,2.859)--(5.122,2.786)--(5.454,2.736)--(5.685,2.697)--(5.786,2.680)%
  --(6.118,2.665)--(6.451,2.654)--(6.783,2.622)--(7.115,2.612)--(7.447,2.583);
\gpcolor{rgb color={0.000,0.000,1.000}}
\draw[gp path,very thick] (3.010,5.631)--(3.129,5.453)--(3.175,5.386)--(3.345,5.142)--(3.461,4.975)%
  --(3.516,4.897)--(3.692,4.653)--(3.793,4.532)--(3.909,4.408)--(4.125,4.184)--(4.145,4.164)%
  --(4.407,3.919)--(4.458,3.876)--(4.697,3.675)--(4.790,3.605)--(5.047,3.430)--(5.122,3.381)%
  --(5.454,3.207)--(5.498,3.186)--(5.786,3.062)--(6.118,2.959)--(6.190,2.941)--(6.451,2.876)%
  --(6.783,2.800)--(7.115,2.749)--(7.447,2.709);
\gpdefrectangularnode{gp plot 1}{\pgfpoint{1.136cm}{0.985cm}}{\pgfpoint{7.447cm}{5.631cm}}
\end{tikzpicture}
}}}
\caption{\label{fig:c} Same as Figure~\ref{fig:b} but for $M=150$
  antennas serving $K=30$ terminals located at random in the cell.
  Here $\tau_p^u=60$, $\tau_p^o=10$ and $M'=7$. The O-terminals were
  located on the cell border, with no additional shadow margin.}
\end{figure}
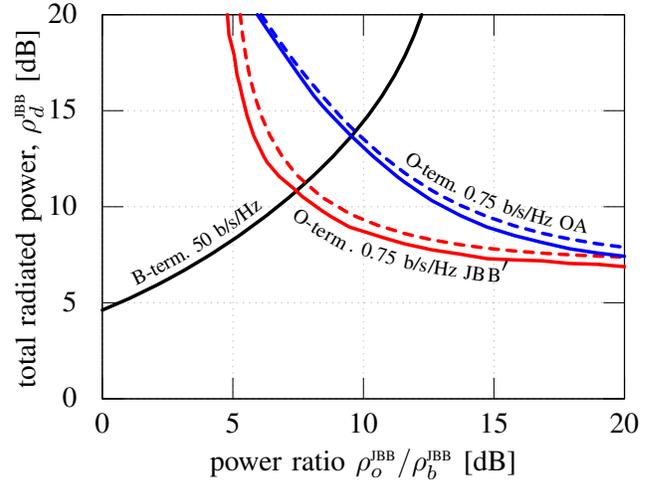

\begin{figure}[t!]
{\centerline{\scalebox{1.1}{\begin{tikzpicture}[gnuplot]
\path (0.000,0.000) rectangle (8.000,6.000);
\gpfill{rgb color={1.000,1.000,1.000}} (1.320,0.985)--(7.446,0.985)--(7.446,5.630)--(1.320,5.630)--cycle;
\gpcolor{color=gp lt color border}
\gpsetlinetype{gp lt border}
\gpsetlinewidth{1.00}
\draw[gp path] (1.320,0.985)--(1.320,5.630)--(7.446,5.630)--(7.446,0.985)--cycle;
\gpcolor{color=gp lt color axes}
\gpsetlinetype{gp lt axes}
\gpsetlinewidth{0.50}
\draw[gp path] (1.320,0.985)--(7.447,0.985);
\gpcolor{color=gp lt color border}
\gpsetlinetype{gp lt border}
\draw[gp path] (1.320,0.985)--(1.571,0.985);
\draw[gp path] (7.447,0.985)--(7.196,0.985);
\gpcolor{rgb color={0.000,0.000,0.000}}
\node[gp node right,font={\fontsize{10pt}{12pt}\selectfont}] at (1.136,0.985) {-10};
\gpcolor{color=gp lt color axes}
\gpsetlinetype{gp lt axes}
\draw[gp path] (1.320,2.147)--(7.447,2.147);
\gpcolor{color=gp lt color border}
\gpsetlinetype{gp lt border}
\draw[gp path] (1.320,2.147)--(1.571,2.147);
\draw[gp path] (7.447,2.147)--(7.196,2.147);
\gpcolor{rgb color={0.000,0.000,0.000}}
\node[gp node right,font={\fontsize{10pt}{12pt}\selectfont}] at (1.136,2.147) {-5};
\gpcolor{color=gp lt color axes}
\gpsetlinetype{gp lt axes}
\draw[gp path] (1.320,3.308)--(7.447,3.308);
\gpcolor{color=gp lt color border}
\gpsetlinetype{gp lt border}
\draw[gp path] (1.320,3.308)--(1.571,3.308);
\draw[gp path] (7.447,3.308)--(7.196,3.308);
\gpcolor{rgb color={0.000,0.000,0.000}}
\node[gp node right,font={\fontsize{10pt}{12pt}\selectfont}] at (1.136,3.308) {0};
\gpcolor{color=gp lt color axes}
\gpsetlinetype{gp lt axes}
\draw[gp path] (1.320,4.470)--(7.447,4.470);
\gpcolor{color=gp lt color border}
\gpsetlinetype{gp lt border}
\draw[gp path] (1.320,4.470)--(1.571,4.470);
\draw[gp path] (7.447,4.470)--(7.196,4.470);
\gpcolor{rgb color={0.000,0.000,0.000}}
\node[gp node right,font={\fontsize{10pt}{12pt}\selectfont}] at (1.136,4.470) {5};
\gpcolor{color=gp lt color axes}
\gpsetlinetype{gp lt axes}
\draw[gp path] (1.320,5.631)--(7.447,5.631);
\gpcolor{color=gp lt color border}
\gpsetlinetype{gp lt border}
\draw[gp path] (1.320,5.631)--(1.571,5.631);
\draw[gp path] (7.447,5.631)--(7.196,5.631);
\gpcolor{rgb color={0.000,0.000,0.000}}
\node[gp node right,font={\fontsize{10pt}{12pt}\selectfont}] at (1.136,5.631) {10};
\gpcolor{color=gp lt color axes}
\gpsetlinetype{gp lt axes}
\draw[gp path] (1.320,0.985)--(1.320,5.631);
\gpcolor{color=gp lt color border}
\gpsetlinetype{gp lt border}
\draw[gp path] (1.320,0.985)--(1.320,1.236);
\draw[gp path] (1.320,5.631)--(1.320,5.380);
\gpcolor{rgb color={0.000,0.000,0.000}}
\node[gp node center,font={\fontsize{10pt}{12pt}\selectfont}] at (1.320,0.677) {0};
\gpcolor{color=gp lt color axes}
\gpsetlinetype{gp lt axes}
\draw[gp path] (2.852,0.985)--(2.852,5.631);
\gpcolor{color=gp lt color border}
\gpsetlinetype{gp lt border}
\draw[gp path] (2.852,0.985)--(2.852,1.236);
\draw[gp path] (2.852,5.631)--(2.852,5.380);
\gpcolor{rgb color={0.000,0.000,0.000}}
\node[gp node center,font={\fontsize{10pt}{12pt}\selectfont}] at (2.852,0.677) {5};
\gpcolor{color=gp lt color axes}
\gpsetlinetype{gp lt axes}
\draw[gp path] (4.384,0.985)--(4.384,5.631);
\gpcolor{color=gp lt color border}
\gpsetlinetype{gp lt border}
\draw[gp path] (4.384,0.985)--(4.384,1.236);
\draw[gp path] (4.384,5.631)--(4.384,5.380);
\gpcolor{rgb color={0.000,0.000,0.000}}
\node[gp node center,font={\fontsize{10pt}{12pt}\selectfont}] at (4.384,0.677) {10};
\gpcolor{color=gp lt color axes}
\gpsetlinetype{gp lt axes}
\draw[gp path] (5.915,0.985)--(5.915,5.631);
\gpcolor{color=gp lt color border}
\gpsetlinetype{gp lt border}
\draw[gp path] (5.915,0.985)--(5.915,1.236);
\draw[gp path] (5.915,5.631)--(5.915,5.380);
\gpcolor{rgb color={0.000,0.000,0.000}}
\node[gp node center,font={\fontsize{10pt}{12pt}\selectfont}] at (5.915,0.677) {15};
\gpcolor{color=gp lt color axes}
\gpsetlinetype{gp lt axes}
\draw[gp path] (7.447,0.985)--(7.447,5.631);
\gpcolor{color=gp lt color border}
\gpsetlinetype{gp lt border}
\draw[gp path] (7.447,0.985)--(7.447,1.236);
\draw[gp path] (7.447,5.631)--(7.447,5.380);
\gpcolor{rgb color={0.000,0.000,0.000}}
\node[gp node center,font={\fontsize{10pt}{12pt}\selectfont}] at (7.447,0.677) {20};
\gpcolor{color=gp lt color border}
\draw[gp path] (1.320,5.631)--(1.320,0.985)--(7.447,0.985)--(7.447,5.631)--cycle;
\gpcolor{rgb color={0.000,0.000,0.000}}
\node[gp node center,rotate=90,font={\fontsize{10pt}{12pt}\selectfont}] at (0.246,3.308) {total radiated power, $\rho_d^\jbb$ [dB]};
\node[gp node center,font={\fontsize{10pt}{12pt}\selectfont}] at (4.383,0.215) {power ratio $\rho_o^\jbb/\rho_b^\jbb$ [dB]};
\gpsetlinetype{gp lt plot 0}
\draw[gp path,very thick,postaction={decorate,decoration={text along path,reverse path,text align={left indent=2mm}, text=|\scriptsize|B-term. 2 b/s/Hz,text align=left,raise=-7pt}}] (5.198,5.631)--(5.190,5.615)--(5.061,5.386)--(4.901,5.142)--(4.867,5.095)%
  --(4.720,4.897)--(4.545,4.687)--(4.516,4.653)--(4.289,4.408)--(4.222,4.343)--(4.038,4.164)%
  --(3.900,4.039)--(3.763,3.919)--(3.577,3.765)--(3.464,3.675)--(3.255,3.516)--(3.137,3.430)%
  --(2.932,3.287)--(2.780,3.186)--(2.610,3.077)--(2.384,2.941)--(2.287,2.885)--(1.965,2.711)%
  --(1.937,2.697)--(1.642,2.554)--(1.407,2.452)--(1.320,2.415);
\gpsetlinetype{gp lt plot 1}
\draw[gp path,very thick] (5.076,5.631)--(4.937,5.386)--(4.867,5.279)--(4.777,5.142)--(4.594,4.897)%
  --(4.545,4.838)--(4.388,4.653)--(4.222,4.475)--(4.159,4.408)--(3.906,4.164)--(3.900,4.158)%
  --(3.628,3.919)--(3.577,3.878)--(3.323,3.675)--(3.255,3.623)--(2.990,3.430)--(2.932,3.390)%
  --(2.624,3.186)--(2.610,3.177)--(2.287,2.983)--(2.212,2.941)--(1.965,2.807)--(1.740,2.697)%
  --(1.642,2.649)--(1.320,2.509);
\gpcolor{rgb color={1.000,0.000,0.000}}
\draw[gp path, very thick] (2.652,5.631)--(2.674,5.386)--(2.703,5.142)--(2.739,4.897)--(2.787,4.653)%
  --(2.849,4.408)--(2.930,4.164)--(2.932,4.159)--(3.035,3.919)--(3.176,3.675)--(3.255,3.567)%
  --(3.367,3.430)--(3.577,3.232)--(3.636,3.186)--(3.900,3.015)--(4.042,2.941)--(4.222,2.861)%
  --(4.545,2.749)--(4.742,2.697)--(4.867,2.665)--(5.190,2.602)--(5.512,2.554)--(5.835,2.517)%
  --(6.157,2.488)--(6.480,2.466)--(6.750,2.452)--(6.802,2.449)--(7.125,2.435)--(7.447,2.425);
\gpcolor{rgb color={0.000,0.000,1.000}}
\draw[gp path,very thick,postaction={decorate,decoration={text along path, text align={left indent=25mm},text=|\scriptsize|O-term. 0.75 b/s/Hz OA,text align=left,raise=3pt}}] (2.999,5.631)--(3.134,5.386)--(3.255,5.166)--(3.268,5.142)--(3.410,4.897)%
  --(3.562,4.653)--(3.577,4.628)--(3.719,4.408)--(3.894,4.164)--(3.900,4.156)--(4.088,3.919)%
  --(4.222,3.768)--(4.313,3.675)--(4.545,3.461)--(4.581,3.430)--(4.867,3.221)--(4.922,3.186)%
  --(5.190,3.035)--(5.392,2.941)--(5.512,2.891)--(5.835,2.780)--(6.144,2.697)--(6.157,2.693)%
  --(6.480,2.626)--(6.802,2.574)--(7.125,2.533)--(7.447,2.501);
\gpcolor{rgb color={1.000,0.000,0.000}}
\gpsetlinetype{gp lt plot 0}
\draw[gp path,very thick,postaction={decorate,decoration={text along path, text align={left indent=30mm},text=|\scriptsize|O-term. 0.75 b/s/Hz JBB$'$\ ,text align=left,raise=-8pt}}] (2.499,5.631)--(2.529,5.386)--(2.545,5.142)--(2.563,4.897)--(2.609,4.653)%
  --(2.610,4.646)--(2.662,4.408)--(2.766,4.164)--(2.848,3.919)--(2.932,3.736)--(2.968,3.675)%
  --(3.149,3.430)--(3.255,3.331)--(3.416,3.186)--(3.577,3.056)--(3.789,2.941)--(3.900,2.873)%
  --(4.222,2.711)--(4.263,2.697)--(4.545,2.615)--(4.867,2.540)--(5.190,2.470)--(5.497,2.452)%
  --(5.512,2.451)--(5.835,2.415)--(6.157,2.381)--(6.480,2.355)--(6.802,2.341)--(7.125,2.326)%
  --(7.447,2.321);
\gpcolor{rgb color={0.000,0.000,1.000}}
\draw[gp path,very thick] (2.960,5.631)--(3.079,5.386)--(3.214,5.142)--(3.255,5.065)--(3.345,4.897)%
  --(3.492,4.653)--(3.577,4.516)--(3.638,4.408)--(3.811,4.164)--(3.900,4.049)--(4.000,3.919)%
  --(4.180,3.675)--(4.222,3.627)--(4.407,3.430)--(4.545,3.310)--(4.735,3.186)--(4.867,3.101)%
  --(5.122,2.941)--(5.190,2.904)--(5.512,2.760)--(5.715,2.697)--(5.835,2.662)--(6.157,2.572)%
  --(6.480,2.510)--(6.787,2.452)--(6.802,2.449)--(7.125,2.426)--(7.447,2.377);
\gpdefrectangularnode{gp plot 1}{\pgfpoint{1.320cm}{0.985cm}}{\pgfpoint{7.447cm}{5.631cm}}
\end{tikzpicture}
}}}
\caption{\label{fig:e1} Same as Figure~\ref{fig:a} but here the
  O-terminals were located halfway between the base station and the
  cell border.}
\end{figure}
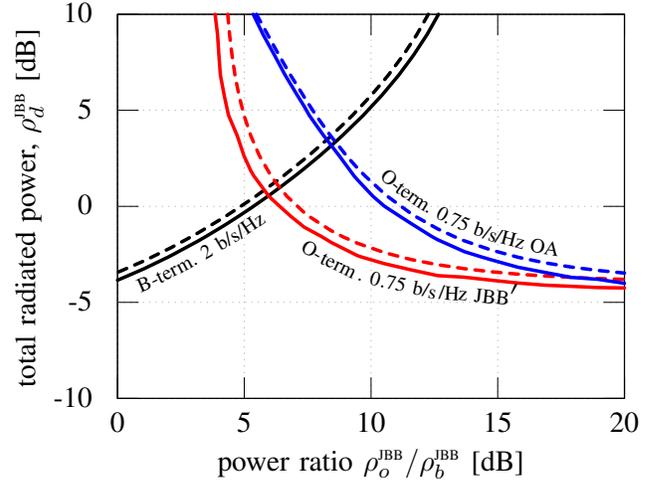
 
\section{Conclusions}
 
The surplus of  spatial degrees of freedom in massive MIMO makes it
possible to ``hide'' signals in the channel nullspace, which terminals
targeted by beamforming do not see. With joint beamforming and
broadcasting (JBB), this opportunity is used to broadcast public information,
aimed at terminals to which the base station does not have
channel state information.   Depending on the selected
operating point, JBB can offer  savings in radiated power  in the order of
3 dB compared to orthogonal access.
An additional, less obvious advantage of JBB is that the broadcast information is
spread over all time-frequency resources, so the maximum possible
time and frequency diversity is always exploited.

\appendices

\section{Calculation of $\E{ \left| \sqrt{\rho_o}\cdot \bg_k^H  \Pi^\perp_{\hat\bG} \bU\bq(t) \right|^2}$}

First note that
\begin{align}
\E{\bg_k\bg_k^H | \hat\bG} = \hat\bg_k\hat\bg_k^H + (\beta_k-\gamma_k)\bI.
\end{align}
Therefore,   since $\Pi^\perp_{\hat\bG}\hat\bg_k=\bzero$,
  \begin{align}
  & 
\ \ \ \E{ \left| \sqrt{\rho_o}\cdot \bg_k^H  \Pi^\perp_{\hat\bG} \bU\bq(t) \right|^2} \nonumber \\
 & = \rho_o \cdot \E{ \left| \bg_k^H  \Pi^\perp_{\hat\bG} \bU\bq(t) \right|^2} \nonumber \\
  & = \frac{\rho_o}{M'}  \cdot \E{ \tr{ \bU^H  \Pi^\perp_{\hat\bG} \bg_k \bg_k^H  \Pi^\perp_{\hat\bG}\bU }} \nonumber \\
  & = \frac{\rho_o}{M'}  \cdot \tr{ \E{ \E{ \bU^H  \Pi^\perp_{\hat\bG} \bg_k \bg_k^H  \Pi^\perp_{\hat\bG} \bU | \hat\bG}}} \nonumber \\
  & = \frac{\rho_o}{M'}  \cdot \tr{ \E{ \bU^H \Pi^\perp_{\hat\bG} \cdot \E{    \bg_k \bg_k^H | \hat\bG} \cdot \Pi^\perp_{\hat\bG} \bU }}\nonumber \\
  & = \frac{\rho_o}{M'}  \cdot \tr{ \E{     \bU^H  \Pi^\perp_{\hat\bG} (\hat\bg_k\hat\bg_k^H + (\beta_k-\gamma_k)\bI)   \Pi^\perp_{\hat\bG} \bU  }} \nonumber \\
  & = \frac{\rho_o}{M'}  (\beta_k-\gamma_k) \cdot \tr{ \E{   \bU^H  \Pi^\perp_{\hat\bG}    \bU  }} \nonumber \\
  & = \rho_o   (\beta_k-\gamma_k).
  \end{align}
  After the third equality sign, double expectations appear. The inner
  expectation is with respect to all sources of randomness but
  conditioned on $\hat\bG$, and the other expectation is with respect
  to the remaining randomness (that is, $\hat\bG$).

\section*{Acknowledgement}

The authors thank the reviewers for their constructive comments, which helped improve the quality of the paper.

 \end{document}